\documentclass[12pt]{iopart}
\usepackage{iopams}
\usepackage{graphicx}
\usepackage{cite}
\usepackage{makecell}

\renewcommand{\vec}[1]{\boldsymbol{#1}}
\renewcommand{\Re}[1]{{\rm Re}\left[#1\right]}
\renewcommand{\Im}[1]{{\rm Im}\left[#1\right]}

\newcommand{\eqref}[1]{(\ref{#1})}
\newcommand{\pd}[2]{\frac{{\partial #1}}{{\partial #2}}}

\def\e{\varepsilon}
\def\t{\theta}

\def\a{\alpha}
\def\b{\beta}
\def\vp{\varphi}
\def\w{\omega}

\begin{document}

\title{High-Order Phase Reduction for Coupled Oscillators}
\author{Erik Gengel$^1$, Erik Teichmann$^1$, Michael Rosenblum$^1$,
Arkady Pikovsky$^{1,2}$}
\address{$^1$ Institute of Physics and Astronomy, University of Potsdam,
Karl-Liebknecht-Str. 24/25, 14476 Potsdam-Golm, Germany}
\address{$^2$ Control Theory Department, Institute of Information Technologies,
Mathematics and Mechanics, Lobachevsky University, Nizhny Novgorod, Russia}

\begin{abstract}
We explore the phase reduction in networks of coupled
oscillators in the higher orders of the coupling parameter. For coupled Stuart-Landau oscillators,
where the phase can be introduced explicitly, we develop an analytic
perturbation procedure to allow for the obtaining of the higher-order approximation explicitly.
We demonstrate this by deriving the second-order phase
equations for a network of three Stuart-Landau oscillators. For systems where
explicit expressions of the phase are not available, we present a numerical
procedure that constructs the phase dynamics equations for a small network of coupled units.
We apply this approach to a network of three van der Pol
oscillators and reveal components in the coupling with different scaling in the
interaction strength.
\end{abstract}
\maketitle

\section{Introduction}

Networks of coupled self-sustained oscillators are widely used to describe
complex rhythmical systems in physics~\cite{Nixon_etal-13,Matheny_etal-14}, biology~\cite{pol1928heartbeat,%
ashraf2016synchronization}, and other fields~\cite{strogatz2005theoretical, dorfler2014synchronization}.
In particular, such models are relevant for the description of laser~\cite{Garca-Ojalvo99} 
and nanomechanical~\cite{Matheny_etal-19} oscillator arrays, coupled Josephson 
junctions~\cite{Cawthorne_etal-99} and spin-torque oscillators~\cite{Tiberkevich_etal-09},  
power grids~\cite{Motter-13}, the activity of neuronal populations~\cite{Ulhaas_etal-09}, 
the interaction of different organs within a human body~\cite{Glass-01}, cell assemblies~\cite{Prindle_etal-12}, etc. 

One of the most famous theoretical tools for the analysis of coupled oscillators is the phase 
reduction~\cite{kuramoto1984chemical,pikovsky2003synchronization,Nakao-16,pietras2019network}.
This approach provides a recipe for a description of complex high-dimensional oscillators by a single cyclic variable, phase $\vp$, so that the dynamics of a network of $N$ generally
high-dimensional elements
reduces to a set of $N$ coupled one-dimensional differential equations for the phases. This reduction is based on the slaving principle (the corresponding notion in mathematical literature is the normal hyperbolicity): While the phases correspond to the neutral directions with zero Lyapunov exponents, the ``amplitudes'' (i.e. all other variables except for the phases) are stable and thus follow the phase dynamics.
As a result, the full system's dynamics reduces to that on the $N$-dimensional torus spanned by the phases.

This reduction allows studying general behaviours of coupled oscillators through
phase dynamics models, a prominent example here is the analytically tractable Kuramoto model of all-to-all interconnected units, derived for a population of weakly coupled oscillators close to the Hopf bifurcation point~\cite{Kuramoto-75,kuramoto1984chemical,Acebron-etal-05}.
Generally, phase dynamics models are expected to be valid
for arbitrary oscillators and weak to moderate coupling. However, the existing theory employs a perturbative approach and provides phase equations only in the first-order approximation in the coupling strength. Derivation of high-order corrections would extend the validity
of the approach beyond the weak-coupling limit and, thus, essentially increase our ability to analyse complex networks. Yet, for the moment it remains a theoretical challenge, in spite of numerous attempts~\cite{leon2019phase,wilson2016isostable,daido1992order,rosenblum2007self,kurebayashi2013phase, pyragas2015phase}.

In addition to obvious advantages for theoretical studies, phase reduction
provides a framework for model reconstruction from measurements
and, hence, plays an essential role in experimental research.
Reconstruction of the phase dynamics equations from data is much more simple than
recovery of equations for the state variables because
(i) phase equations are low-dimensional and (ii) have a simple universal form
because their right-hand sides are $2\pi$-periodic functions of the phases.
Though precise estimation of the phases from scalar signals remains a topic
of ongoing studies \cite{gengel2019phase,rosenblum2018inferring,kralemann2008phase}, phase models have been successfully recovered for several laboratory experiments as well as for physiological systems
\cite{Bezruchko2003,Tokuda-Jain-Kiss-Hudson-07,Blaha-etal-11,Kralemann_etal-13,%
Rosenblum-Fruehwirth-Moser-Pikovsky-19,ticcinelli2017coherence,%
topccu2018disentangling,stankovski2017coupling,stankovski2016alterations}.
In particular, this approach yields a way to tackle the connectivity problem, i.e. to recover directional causal links between oscillatory
sources solely from multivariate data.

However, for more than two oscillators and moderate coupling
the connectivity, as it appears in phase equations, generally differs from
the connectivity as defined by physical connections, e.g. the units that
are not directly coupled appear as interacting with each other or
the interaction of pairwise physically connected elements appears
as non-pairwise in the phase description. These effects are due to
the terms that do not appear in the first-order phase approximation.
Thus, reconstruction and interpretation of phase models requires better
understanding of the high-order phase reduction \cite{Matheny_etal-19,kralemann2011reconstructing,kralemann2014reconstructing}.

In this paper, we take a step to explore high-order phase dynamics. Our main goal
is to match theoretical derivation of higher-order phase coupling terms with the data analysis.
After a brief introduction to phase reduction approach in Section~\ref{sec coup lim and pr},
in the rest of Section~\ref{sec analysis}
we outline the perturbation approach and apply it to a particular example
of three coupled Stuart-Landau oscillators. Our technique is close to the procedure of Ref.~\cite{leon2019phase},
but is better suited for generic networks and different natural frequencies of the oscillators.
For this system, we derive and discuss phase dynamics equations in
the higher orders in the coupling parameter
(we present the second-order terms in full details, while for the higher orders 
we discuss only the phase dependence of the terms).
Next, in Section~\ref{sec methods} we present a numerical framework
for the reconstruction of the phase equations from the simulations.
First, we apply this framework to the
Stuart-Landau network and in this way verify our analytical results. Second, we analyse a network of three van der Pol oscillators for which only a numerical study is currently possible,
and identify the scaling of different coupling terms with the coupling strength parameter.
In Section~\ref{sec:concl} we discuss our results.

\section{Theoretical analysis} \label{sec analysis}
\subsection{Coupled oscillators and first-order phase reduction}\label{sec coup lim and pr}
We first briefly recall the basic ideas of the phase reduction in the first order in the coupling
parameter $\varepsilon$ (see Refs.~\cite{pietras2019network,pikovsky2003synchronization} for more details).
One starts by considering an oscillatory dynamical system
\[
\frac{d\mathbf{y}}{dt}=\mathbf{f}(\mathbf{y})
\]
possessing  a stable limit cycle $\mathbf{Y}(t)=\mathbf{Y}(t+T)$ in its state space.
As is well-known, on the limit cycle and in its basin of attraction one can define
the phase $\varphi=\Phi(\mathbf{y})$ which grows uniformly in time
\[
\dot\varphi=\omega=\frac{\partial \Phi}{\partial \mathbf{y}}\mathbf{f}(\mathbf{y})
\]
with basic frequency $\omega=2\pi/T$. Notice,
that on the limit cycle the system's state is uniquely defined by the phase:
$\mathbf{y}=\mathbf{Y}(\varphi)$.
Outside of the limit cycle, this is not true: one also has to know the deviation
from the cycle.

In the context of oscillatory networks, one considers many coupled oscillators
that we label by index $k$:
\begin{equation}
\frac{d\mathbf{y}_k}{dt}=\mathbf{f}_k(\mathbf{y}_k)+
\varepsilon \mathbf{G}_k(\mathbf{y}_1,\mathbf{y}_2,\ldots)\;.
\label{eq:gen_network}
\end{equation}
Here $\e$ is the small parameter governing the strength of the coupling.
The equation for the phases is obtained by exploiting their definition $\varphi_k=\Phi_k(\mathbf{y}_k)$:
\begin{equation}
\dot\varphi_k=\omega_k+\varepsilon\frac{\partial \Phi_k}{\partial \mathbf{y}_k}\mathbf{G}_k(\mathbf{y}_1,\mathbf{y}_2,\ldots)\;.
\label{eq:pertlc}
\end{equation}
This equation is exact, but it contains the full state space trajectories $\{\mathbf{y}_j\}$.
However, for a small perturbation these trajectories are close to the limit cycle, up
to deviations of order $\sim \varepsilon$. Thus, substituting the zero-order
approximation $\mathbf{y}_j\approx \mathbf{Y}_j$ into the r.h.s.
of \eqref{eq:pertlc}, we obtain equations, where only the states on the
limit cycles appear, and these states are unambiguously determined by the phases:
\[
\dot\varphi_k=\omega_k+\varepsilon\frac{\partial \Phi_k}{\partial \mathbf{Y}_k}\mathbf{G}_k(\mathbf{Y}_1,\mathbf{Y}_2,\ldots)=\omega_k+\varepsilon G_k(\varphi_k,\varphi_1,\varphi_2,\ldots)\;.
\]

As is clear from the presented consideration, to extend the reduction beyond the first-order approximation,
one has to calculate the deviations from the limit cycle of orders $\sim\varepsilon,\varepsilon^2,\ldots$
and to express these deviations in terms of the phases.
Below, we do not provide a general approach valid for arbitrary systems, but
restrict ourselves to the simplest case where the phase can be introduced analytically.

\subsection{A network of Stuart-Landau oscillators}
It is instructive to introduce the Stuart-Landau oscillators in dimensional variables, to
understand the meaning of dimensionless parameters in the problem. We write
the equation for the complex amplitude of oscillations $Z$ for a particular oscillator as
\begin{equation}
\frac{d Z}{d\tau}=(\mu+i\nu)Z-(\beta+i\gamma)|Z|^2Z+\epsilon G(Z_1,Z_2,\ldots)\;,
\label{eq:z}
\end{equation}
where $G$ is the coupling term depending on the states of other elements of the network.
Notice that for brevity and without loss of generality we omit the index for the considered
oscillator and label other units by the index $k=1,2,\ldots$.

In Eq.~\eqref{eq:z}, all the parameters are dimensional, and it is convenient to
perform a transformation to dimensionless variables and parameters. Here one should also
take into account, that generally the parameters might be different for different oscillators.
Using only local parameters, one can introduce
the dimensionless amplitude $Z=\sqrt{\mu/\beta}A$, so that all uncoupled oscillators
will have amplitude one. Another variable that we can scale is the time, and here
one has to choose some common scaling for all oscillators. It appears convenient to assume that the growth rate of linear oscillations $\mu$ is the same for all
oscillators  (the results can be straightforwardly generalized also to the case of different $\mu$), and use it to introduce dimensionless time as
$t = \mu \tau$. Then we obtain
\begin{equation}
\dot A=(1+i\omega)A-|A|^2A-i\alpha A(|A|^2-1)+\varepsilon G(A_1, A_2,\ldots)\;,
\label{eq:1sl}
\end{equation}
where $\dot A$ now means derivative with respect to $t$.
Here $\omega=\nu/\mu-\gamma/\beta$ is the dimensionless
frequency of the limit cycle oscillation
and the limit cycle has unit amplitude $|A^{(0)}|=1$. Parameter $\alpha=\gamma/\beta$
measures non-isochronicity, as we will see below, it determines the
phase definition function $\Phi(A)$. The dimensionless
coupling parameter $\varepsilon$ depends
on the scaling of the function $G$. If $G$ contains first powers of the amplitudes $Z$, then
$\varepsilon=\epsilon/\mu$, where we assume
that $\mu/\beta \approx \mu_k/\beta_k$, i.e. all the coupled oscillators $Z$
have similar amplitudes.
As we see, all the dimensionless parameters entering the
problem, namely $\omega,\alpha,\varepsilon$, are the original parameters of the system
normalized by the linear growth rate of oscillations $\mu$.

For further analysis it is instructive to write the Stuart-Landau equations in polar coordinates, with
the amplitude $R$ and the angle $\theta$, so that $A=Re^{i\theta}$ and
\begin{eqnarray}
\dot R&=&R-R^3+\varepsilon \Re{e^{-i\theta}G}\;,
\label{eq:rth1}\\
\dot\theta&=&\omega-\alpha(R^2-1)+\e R^{-1}
\Im{e^{-i\theta}G}\;.
\label{eq:rth2}
\end{eqnarray}
It is straightforward to check that, in the absence of coupling, the quantity
$\varphi=\theta-\alpha\ln R$ grows uniformly in time and, hence, is the true phase.
Therefore, we rewrite Eqs.~(\ref{eq:rth1},\ref{eq:rth2})
in terms of $R,\varphi$:
\begin{eqnarray}
\dot R&=&R-R^3+\varepsilon \Re{e^{-i(\varphi+\alpha\ln R)}
G(R_1,\varphi_1,R_2,\varphi_2,\ldots)}\label{eq:rph1}\;,\\
\dot\varphi&=&\omega+\frac{\varepsilon}{R}[
\Im{e^{-i(\varphi+\alpha\ln R)}G(R_1,\varphi_1,R_2,\varphi_2,\ldots)}\nonumber\\
&&-\alpha \Re{e^{-i(\varphi+\alpha\ln R)}G(R_1,\varphi_1,R_2,\varphi_2,\ldots)}\;.
\label{eq:rph2}
\end{eqnarray}
Here we have explicitly used that the coupling terms depend on the amplitudes
and the phases of other oscillators.

\subsection{Outline of the perturbation method}

Now, we outline the exploited perturbation procedure, while in the next subsection we
will elaborate on it for a particular example. The idea is to look for a dynamics, in which
the amplitudes are enslaved by the phases. Namely, we assume that the amplitude $R$
is a function of the phases:
\begin{equation}
R=1+\varepsilon r^{(1)}(\varphi,\varphi_1,\varphi_2,\ldots)+
\varepsilon^2 r^{(2)}(\varphi,\varphi_1,\varphi_2,\ldots)+\ldots\,,
\label{eq:R}
\end{equation}
and that the dynamics of the phases is also represented as a power series in $\varepsilon$:
\begin{equation}
\dot\varphi=\omega+\varepsilon\Psi^{(1)}(\varphi,\varphi_1,\varphi_2,\ldots)+\varepsilon^2\Psi^{(2)}(\varphi,\varphi_1,\varphi_2,\ldots)+\ldots\,.
\label{eq:Q}
\end{equation}
We substitute these expressions, together with the similar expressions for $R_k$,
$\varphi_k$, into Eqs.~(\ref{eq:rph1},\ref{eq:rph2}). Equating the terms
in each power of $\varepsilon$, we obtain the unknown functions $r^{(1)},r^{(2)},\Psi^{(1)},\Psi^{(2)},\ldots$.

We illustrate here the first few steps.
In the equation for the phase, the first-order approximation, as discussed above, corresponds
to taking into account only the leading term in Eq.~\eqref{eq:R}, this yields
\[
\Psi^{(1)}=\Im{e^{-i\varphi}G(1,\varphi_1,1,\varphi_2,\ldots)} -\alpha \Re{e^{-i\varphi}G(1,\varphi_1,1,\varphi_2,\ldots)}\;.
\]
The equation for the amplitude in the first order is obtained by substituting
Eq.~\eqref{eq:R} in Eq.~\eqref{eq:rph1}:
\[
\frac{dr^{(1)}}{dt}=-2 r^{(1)}+ \Re{e^{-i\varphi}
G(1,\varphi_1,1,\varphi_2,\ldots)}\;.
\]
We express the time derivative of $r^{(1)}$ via partial derivatives with respect to the phases,
and insert the zero-order expressions for these derivatives:
\begin{eqnarray}
\frac{dr^{(1)}}{dt}&=&\frac{\partial r^{(1)}}{\partial \varphi}\dot\varphi+ \frac{\partial r^{(1)}}{\partial \varphi_1}\dot\varphi_1+
 \frac{\partial r^{(1)}}{\partial \varphi_2}\dot\varphi_2+\ldots\approx\nonumber\\&&
 \approx \frac{\partial r^{(1)}}{\partial \varphi}\omega+\frac{\partial r^{(1)}}{\partial \varphi_1} \omega_1
 +
 \frac{\partial r^{(1)}}{\partial \varphi_2}\omega_2+\ldots\;.
 \label{eq:pder}
\end{eqnarray}
Thus, the problem of determining first-order correction to the amplitude reduces
to the partial differential equation
\begin{eqnarray}
2 r^{(1)}+\frac{\partial r^{(1)}}{\partial \varphi}\omega+\frac{\partial r^{(1)}}{\partial \varphi_1} \omega_1
 +
 \frac{\partial r^{(1)}}{\partial \varphi_2}\omega_2+\ldots=&&\nonumber\\
 \Re {e^{-i\varphi}
G(1,\varphi_1,1,\varphi_2,\ldots)}=\sum_{m,m_1,m_2,\ldots}g_{mm_1m_2\ldots}
e^{i(m\varphi+m_1\varphi_1+m_2\varphi_2+\ldots)}&&\label{eq:pde}
\end{eqnarray}
Here, we used that the function $G$ is $2\pi$-periodic with respect to the phases and, hence, the r.h.s. can be
written as a multiple Fourier series. Similarly, we expand the function $r^{(1)}$ as:
\begin{equation*}
r^{(1)}=\sum_{m,m_1,m_2,\ldots} \rho_{mm_1m_2\ldots} e^{i(m\varphi+m_1\varphi_1+m_2\varphi_2+\ldots)}\;,
\end{equation*}
which finally yields an expression for the Fourier coefficients of $r^{(1)}$
\begin{equation}
\rho_{mm_1m_2\ldots}=\frac{g_{mm_1m_2\ldots}}
{2+im\omega+im_1\omega_1+im_2\omega_2+\ldots}\;.
\label{eq:spde}
\end{equation}
These are the basic steps in the perturbation expansion. The expressions for $r^{(1)},r_k^{(1)}$,
being substituted in Eq.~\eqref{eq:rph2} provide phase equations in order $\sim \e^2$. Equations
for $r^{(2)}$ are partial differential equations of type  \eqref{eq:pde} with r.h.s containing
also $\Psi^{(1)},\;r^{(1)}$, etc.

\subsection{Example: Three coupled Stuart-Landau oscillators}
\label{sec:3sltheory}

In this Section, we exemplify the perturbative procedure for the derivation of phase dynamics
equations in a higher order of the perturbation parameter $\varepsilon$
with a particular configuration of three coupled Stuart-Landau oscillators.

\subsubsection{Configuration of the network.}

We consider an array of three elements with the coupling structure $1\leftrightarrow 2\leftrightarrow 3$
and write the equations in the form~\eqref{eq:1sl}:
\begin{equation}
\fl
\eqalign{
\qquad\dot A_1&=(1+\rmi\omega_1)A_1-|A_1|^2A_1-\rmi \alpha A_1(|A_1|^2-1)+\varepsilon c_{2,1}e^{\rmi\beta_{2,1}} A_2\;,\\
\qquad\dot A_2&=(1+\rmi\omega_2)A_2-|A_2|^2A_2-\rmi\alpha A_2 (|A_2|^2-1)+
\varepsilon \left(c_{1,2}e^{\rmi\beta_{1,2}} A_1 +c_{3,2}e^{\rmi\beta_{3,2}} A_3\right)\;,\\
\qquad\dot A_3&=(1+\rmi\omega_3)A_3-|A_3|^2A_3-\rmi\alpha A_3 (|A_3|^2-1)+\varepsilon c_{2,3}e^{\rmi\beta_{2,3}} A_2\;.
\label{eq:LS3}
}
\end{equation}
Notice that the oscillators have different frequencies. Introducing the amplitudes and
the angle variables according to $A_k=R_k\exp[\rmi\theta_k]$ and the phases $\varphi_k=\theta_k-\alpha
\ln R_k$, we obtain a system of equations in the form~(\ref{eq:rph1}-\ref{eq:rph2}):
\begin{equation}
\eqalign{
\dot R_1&=R_1-R_1^3+\varepsilon c_{2,1}R_2\cos(\theta_2-\theta_1+\beta_{2,1})\;,\\
\dot R_2&=R_2-R_2^3+\varepsilon c_{1,2}R_1\cos(\theta_1-\theta_2+\beta_{1,2})\\
&+\varepsilon c_{3,2}R_3\cos(\theta_3-\theta_2+\beta_{3,2})\;,\\
\dot R_3&=R_3-R_3^3+\varepsilon c_{2,3}R_2\cos(\theta_2-\theta_3+\beta_{2,3})\;,\\
\dot{\vp_1} &= \omega_1 + \e c_{2,1}\frac{R_2}{R_1}[\sin(\t_2-\t_1+\beta_{2,1}) -\a \cos(\t_2-\t_1+\beta_{2,1})]\;, \\
\dot{\vp_2} &= \omega_2 + \e c_{1,2}\frac{R_1}{R_2}[\sin(\t_1-\t_2+\beta_{1,2}) -\a \cos(\t_1-\t_2+\beta_{1,2})] \\
 & + \e c_{3,2}\frac{R_3}{R_2}[\sin(\t_3-\t_2+\beta_{3,2}) -\a \cos(\t_3-\t_2+\beta_{3,2})]\;, \\
\dot{\vp_3} &= \omega_3 + \e c_{2,3}\frac{R_2}{R_3}[\sin(\t_2-\t_3+\beta_{2,3}) -\a \cos(\t_2-\t_3+\beta_{2,3})]\;.
}
\label{eq:3Rph}
\end{equation}

\subsubsection{Small parameter expansion.}

Now we expand the amplitude deviations in powers of $\e$ (below $k=1,2,3$):
\begin{equation}
R_k=1+\e r_k^{(1)}+\e^2 r_k^{(2)}+\e^3r_k^{(3)}+\ldots\;.
\label{eq:rexp}
\end{equation}
According to this, the angles $\t_k$ can be represented as
\begin{equation}
\t_k = \vp_k + \a \left[\e r_k^{(1)} + \e^2 r_k^{(2)}
- 0.5\e^2\left ({r_k^{(1)}}\right )^2\right]+\ldots\;.
\label{eq:texp}
\end{equation}
Also the ratios of the amplitudes, entering \eqref{eq:3Rph}, can be expressed as
\begin{equation}
\frac{R_m}{R_k} = 1+\e\left[r_m^{(1)}-r_k^{(1)}\right]+\e^2
\left[r_m^{(2)}-r_k^{(2)}-r_m^{(1)}r_k^{(1)}+\left(r_k^{(1)}\right)^2\right]+\ldots\;.
\label{eq:Rrat}
\end{equation}
Substituting these expansions in Eq.~\eqref{eq:3Rph}, we obtain the following expressions for the dynamics
of the phases, up to the order $\e^3$:
\begin{equation}
\eqalign{
\dot\vp_1&=\omega_1 + \e c_{2,1}[\sin(\vp_2-\vp_1+\beta_{2,1}) -
\a \cos(\vp_2-\vp_1+\beta_{2,1})] \\
& + \e^2 c_{2,1}(1+\a^2)\sin(\vp_2-\vp_1+\beta_{2,1})(r_2^{(1)}-r_1^{(1)})\\
&+\e^3 c_{2,1}(1+\a^2)\Big[\left(r_2^{(2)}-r_1^{(2)}-r_2^{(1)}r_1^{(1)}+(r_1^{(1)})^2\right)
\sin(\vp_2-\vp_1+\beta_{2,1})\\
  &+\a \left(0.5 (r_1^{(1)})^2+0.5(r_2^{(1)})^2-r_2^{(1)}r_1^{(1)}\right)\cos(\vp_2-\vp_1+\beta_{2,1})\Big]+\ldots\;,\\
\dot\vp_2&=\omega_2 + \e c_{1,2}[\sin(\vp_1-\vp_2+\beta_{1,2}) -
\a \cos(\vp_1-\vp_2+\beta_{1,2})] \\
&+\e c_{3,2}[\sin(\vp_3-\vp_2+\beta_{3,2}) -
\a \cos(\vp_3-\vp_2+\beta_{3,2})] \\
& + \e^2 c_{1,2}(1+\a^2)\sin(\vp_1-\vp_2+\beta_{1,2})(r_1^{(1)}-r_2^{(1)})\\
&+ \e^2 c_{3,2}(1+\a^2)\sin(\vp_3-\vp_2+\beta_{3,2})(r_3^{(1)}-r_2^{(1)})\\
&+\e^3 c_{1,2}(1+\a^2)\Big[\left(r_1^{(2)}-r_2^{(2)}-r_1^{(1)}r_2^{(1)}+(r_2^{(1)})^2\right)
\sin(\vp_1-\vp_2+\beta_{1,2})\\
  &+\a \left(0.5 (r_2^{(1)})^2+0.5(r_1^{(1)})^2-r_1^{(1)}r_2^{(1)}\right)\cos(\vp_1-\vp_2+\beta_{1,2})\Big]\\
&+\e^3 c_{3,2}(1+\a^2)\Big[\left(r_3^{(2)}-r_2^{(2)}-r_3^{(1)}r_2^{(1)}+(r_2^{(1)})^2\right)
\sin(\vp_3-\vp_2+\beta_{3,2})\\
  &+\a \left(0.5 (r_2^{(1)})^2+0.5(r_3^{(1)})^2-r_3^{(1)}r_2^{(1)}\right)\cos(\vp_3-\vp_2+\beta_{3,2})\Big]+\ldots\;,\\
\dot\vp_3&=\omega_3 + \e c_{2,3}[\sin(\vp_2-\vp_3+\beta_{2,3}) -
\a \cos(\vp_2-\vp_3+\beta_{2,3})] \\
& + \e^2 c_{2,3}(1+\a^2)\sin(\vp_2-\vp_3+\beta_{2,3})(r_2^{(1)}-r_3^{(1)})\\
&+\e^3 c_{2,3}(1+\a^2)\Big[\left(r_2^{(2)}-r_3^{(2)}-r_2^{(1)}r_3^{(1)}+(r_3^{(1)})^2\right)
\sin(\vp_2-\vp_3+\beta_{2,3})\\
  &+\a \left(0.5 (r_3^{(1)})^2+0.5(r_2^{(1)})^2-r_2^{(1)}r_3^{(1)}\right)\cos(\vp_2-\vp_3+\beta_{2,3})\Big]+\ldots\;.
}
\label{eq:ph3ord}
\end{equation}

Next, we have to evaluate corrections $r_k^{(1)},r_k^{(2)},\ldots$. This is accomplished
by substituting expressions~\eqref{eq:rexp} in the equations for the amplitudes in \eqref{eq:3Rph}.
Here the time derivatives are calculated according to the chain rule, as the corrections $r_k^{(1)},r_k^{(2)},\ldots$
are assumed to be functions of the phases $\vp_k$.
For the sake of brevity, we present only the formulas for $r_1$:
\begin{equation}
\eqalign{
&\omega_1\pd{r_1^{(1)}}{\vp_1} + \omega_2\pd{r_1^{(1)}}{\vp_2}  +
\omega_3\pd{r_1^{(1)}}{\vp_3}  +2r_1^{(1)} = c_{2,1}\cos(\vp_2-\vp_1+\beta_{2,1}) \;,\\
&\omega_1\pd{r_1^{(2)}}{\vp_1} + \omega_2\pd{r_1^{(2)}}{\vp_2}  +
\omega_3\pd{r_1^{(2)}}{\vp_3}  +2r_1^{(2)} =
-3(r_1^{(1)})^2\\&-\a c_{2,1}(r_2^{(1)}-r_1^{(1)})\sin(\vp_2-\vp_1+\beta_{2,1})
+c_{2,1}r_2^{(1)}\cos(\vp_2-\vp_1+\beta_{2,1})\\
&+c_{2,1}[\sin(\vp_2-\vp_1+\beta_{2,1})-\a \cos(\vp_2-\vp_1+\beta_{2,1})]\pd{r_1^{(1)}}{\vp_1}\\
&+\Big[c_{1,2}[\sin(\vp_1-\vp_2+\beta_{1,2})-\a \cos(\vp_1-\vp_2+\beta_{1,2})]\\
&+c_{3,2}[\sin(\vp_3-\vp_2+\beta_{3,2})-\a \cos(\vp_3-\vp_2+\beta_{3,2})] \Big] \pd{r_1^{(1)}}{\vp_2}\\
 &+c_{2,3}[\sin(\vp_2-\vp_3+\beta_{2,3})-\a \cos(\vp_2-\vp_3+\beta_{2,3})]\pd{r_1^{(1)}}{\vp_3} \;.
}
\end{equation}
The partial differential equations for $r_k^{(1)},r_k^{(2)},\ldots$ are straightforwardly solved
in the Fourier representation, because the variations of the amplitudes are $2\pi$-periodic functions
of the phases, as outlined in discussion of Eq.~\eqref{eq:spde} above.
As a result, we obtain in the 1st order in $\e$:
\begin{equation}
\fl
\eqalign{
r_1^{(1)}&=\frac{2c_{2,1}}{4+(\w_2-\w_1)^2}\cos(\vp_2-\vp_1+\beta_{2,1})+
\frac{(\w_2-\w_1)c_{2,1}}{4+(\w_2-\w_1)^2}\sin(\vp_2-\vp_1+\beta_{2,1})\;,\\
r_2^{(1)}&=\frac{2c_{1,2}}{4+(\w_1-\w_2)^2}\cos(\vp_1-\vp_2+\beta_{1,2})+
\frac{(\w_1-\w_2)c_{1,2}}{4+(\w_1-\w_2)^2}\sin(\vp_1-\vp_2+\beta_{1,2})\\
&+\frac{2c_{3,2}}{4+(\w_3-\w_2)^2}\cos(\vp_3-\vp_2+\beta_{3,2})+
\frac{(\w_3-\w_2)c_{3,2}}{4+(\w_3-\w_2)^2}\sin(\vp_3-\vp_2+\beta_{3,2})\;,\\
r_3^{(1)}&=\frac{2c_{2,3}}{4+(\w_2-\w_3)^2}\cos(\vp_2-\vp_3+\beta_{2,3})+
\frac{(\w_2-\w_3)c_{2,3}}{4+(\w_2-\w_3)^2}\sin(\vp_2-\vp_3+\beta_{2,3})\;.
}
\label{eq:r1}
\end{equation}

Substitution of these expression in Eq.~\eqref{eq:ph3ord} completes the second-order
phase reduction and yields closed equations:
\begin{equation}
\fl
\eqalign{
\dot\vp_1&=\omega_1 + \e c_{2,1}[\sin(\vp_2-\vp_1+\beta_{2,1}) -
\a \cos(\vp_2-\vp_1+\beta_{2,1})] \\
& + \e^2 \Big[a^{(2)}_{1;0} + a^{(2)}_{1;-2,2,0}\cos(2\vp_2-2\vp_1) + b^{(2)}_{1;-2,2,0}\sin(2\vp_2-2\vp_1)\\
& + a^{(2)}_{1;-1,2,-1}\cos(2\vp_2-\vp_1-\vp_3)+ b^{(2)}_{1;-1,2,-1}\sin(2\vp_2-\vp_1-\vp_3)\\
& + a^{(2)}_{1;-1,0,1}\cos(\vp_3-\vp_1) + b^{(2)}_{1;-1,0,1}\sin(\vp_3-\vp_1)\Big]\;,
}
\label{eq:ph2ord1}
\end{equation}
\begin{equation}
\fl
\eqalign{
\dot\vp_2&=\omega_2 + \e c_{1,2}[\sin(\vp_1-\vp_2+\beta_{1,2}) -
\a \cos(\vp_1-\vp_2+\beta_{1,2})] \\
&+\e c_{3,2}[\sin(\vp_3-\vp_2+\beta_{3,2}) -
\a \cos(\vp_3-\vp_2+\beta_{3,2})] \\
& + \e^2 \Big[ a^{(2)}_{2;0} + a^{(2)}_{2;2,-2,0}\cos(2\vp_1-2\vp_2) + b^{(2)}_{2;2,-2,0}\sin(2\vp_1-2\vp_2)\\
& + a^{(2)}_{2;0,-2,2}\cos(2\vp_3-2\vp_2) + b^{(2)}_{2;0,-2,2}\sin(2\vp_3-2\vp_2)\\
& + a^{(2)}_{2;-1,2,-1}\cos(2\vp_2-\vp_1-\vp_3)+ b^{(2)}_{2;-1,2,-1}\sin(2\vp_2-\vp_1-\vp_3)\\
& + a^{(2)}_{2;1,0,-1}\cos(\vp_1-\vp_3) + b^{(2)}_{2;1,0,-1}\sin(\vp_1-\vp_3)\Big]\;,
}
\label{eq:ph2ord2}
\end{equation}
\begin{equation}
\fl
\eqalign{
\dot\vp_3&=\omega_3 + \e c_{2,3}[\sin(\vp_2-\vp_3+\beta_{2,3}) -
\a \cos(\vp_2-\vp_3+\beta_{2,3})] \\
& + \e^2 \Big[a^{(2)}_{3;0} + a^{(2)}_{3;0,2,-2}\cos(2\vp_2-2\vp_3) + b^{(2)}_{3;0,2,-2}\sin(2\vp_2-2\vp_3)\\
& + a^{(2)}_{3;-1,2,-1}\cos(2\vp_2-\vp_3-\vp_1)+ b^{(2)}_{3;-1,2,-1}\sin(2\vp_2-\vp_3-\vp_1)\\
& + a^{(2)}_{3;1,0,-1}\cos(\vp_1-\vp_3) + b^{(2)}_{3;1,0,-1}\sin(\vp_1-\vp_3)\Big]\;.
}
\label{eq:ph2ord3}
\end{equation}

The coefficients of the second-order coupling terms, denoted in
Eqs.~(\ref{eq:ph2ord1}-\ref{eq:ph2ord3})
by $a_{k;\vec{l}}^{(2)},b_{k;\vec{l}}^{(2)}$, are listed
in Tables~\ref{app tab 1}, \ref{app tab 2}, \ref{app tab 3}). Here the 3-component
vector $\vec{l}=(l_1,l_2,l_3)$ is used to signify the term with the combination of the phases
$l_1\vp_1+l_2\vp_2+l_3\vp_3$. Furthermore, in Table~\ref{tab:homodes} we list the terms (without coupling coefficients) appearing in orders $\e^3,\e^4$.

We now shortly discuss the physical meaning of the terms that appear in higher orders in $\e$.
\begin{enumerate}
\item In the second-order approximation, there are no correction terms to the first-order couplings.
These corrections  appear in the third order (and, presumably, in all odd orders).
\item There are terms, which can be roughly described as ``squares'' of the basic coupling terms;
for the dynamics of the 1st oscillator $\varphi_1$ these are constant terms and terms containing
the second harmonics of the phase difference, e.g. $\sim\sin(2\vp_2-2\vp_1)$.
These high-order terms do not arise from the interaction within the whole network;
they appear already in a system of two coupled oscillators.
\item Terms containing combinations of all
three phases, e.g., $\sim \sin(2\vp_2-\vp_1-\vp_3)$,
mean that an effective hypernetwork with non-pairwise coupling
appears already in the second-order reduction (cf. studies of synchronization on hypernetworks
of phase oscillators~\cite{Komarov-Pikovsky-15b,Gong-Pikovsky-19}).
\item Terms containing phase differences for not directly coupled oscillators,
(e.g., the term $\sim \sin(\vp_3-\vp_1)$ on the r.h.s. of the equation for $\dot\vp_1$)
mean that connections in terms of phase dynamics do not coincide with the ``structural'' connections in
the original formulation.
\item While the first-order coupling terms are frequency-independent, the second-order terms depend
explicitly on frequency differences. In a more general setup
(cf. a system of coupled van der Pol equations treated below) we expect that coupling
will depend on the frequencies themselves.
\end{enumerate}


\section{Numerical phase reduction}\label{sec methods}
Stuart-Landau oscillator represents an exceptional case when the
phase and the instantaneous frequency can be directly obtained
from equations. We exploited this feature to derive
the second-order phase dynamics equation in the previous Section.
For a general oscillator, one has to evaluate the phases numerically.
This immediately provides the first-order approximation of the
phase dynamics via numerically calculated phase sensitivity functions.

In this Section, we describe a numerical procedure for determining
coupling functions in higher orders, by virtue of the phase analysis
of numerically obtained trajectories of the full system.
We will first verify it by comparing the results for the Stuart-Landau
model~(\ref{eq:LS3}) with theory in Section~\ref{sec analysis}, and then apply
it to a network of three interacting van der Pol oscillators.

\subsection{Numerical computation of phases and their derivatives}
\label{app: num phase freq calc}
The first step in numerical analysis is the determination of phases
$\vp_i$ and their derivatives
$\dot\vp_i$ for all elements of the analyzed network, $i=1,\ldots,N$.
For this purpose, we extend the technique suggested
in~\cite{rosenblum2019numerical}, where it was described how to obtain phases
from an arbitrary trajectory.

Consider a particular oscillator within a network described by
Eq.~(\ref{eq:gen_network}).
Omitting the index of this unit for simplicity of presentation,
we write
\begin{equation}
\frac{d\mathbf{y}}{dt}=\mathbf{f}(\mathbf{y})+\e\mathbf{G}\;,
\label{app:eq1}
\end{equation}
where the last term describes the coupling to all other units.
As a preparatory step we compute the autonomous period $T$.
It is, for $\e=0$ we take an arbitrary point on the limit cycle
$\mathbf{Y}$,
and assign to it $\vp=0$; the return time to this point is exactly $T$. Now,
for any other point on the limit cycle we can compute the time
$\tau(\mathbf{Y})$ required to
reach the zero point where $\vp=0$ or, equivalently, $\vp=2\pi$. Since the true phase
grows linearly in time and gains $2\pi$ with one revolution around the cycle,
its value can be obtained as
$\vp(\mathbf{Y})=2\pi\frac{T-\tau(\mathbf{Y})}{T}$.

The next step is to compute $\vp(t)$ and $\dot\vp(t)$ for an arbitrary trajectory,
on which
at some time $t$ the oscillator has the state $\mathbf{u}=\mathbf{y}(t)$, and the time
derivative of this state is $\mathbf{v}=\dot\mathbf{y}(t)$.
To this end, we introduce an autonomous copy of the investigated unit:
\begin{equation}
\frac{d\mathbf{w}}{dt}=\mathbf{f}(\mathbf{w})\;,
\label{app:eq2}
\end{equation}
and let this auxiliary system evolve from initial
conditions $\mathbf{w}(0)=\mathbf{u}$,
for the time interval $nT$, where the integer $n$ shall
be large enough to ensure that the trajectory attracts to the limit cycle.
(Practically, we stop the evolution when
$\|\mathbf{w}\left((n-1)T\right)-\mathbf{w}(nT)\|$
is smaller than a given tolerance.)
Since the time interval of the evolution is a multiple of the period, the
initial point $\mathbf{w}(0)=\mathbf{u}$ and the end point
$\mathbf{w}(nT)=\bar\mathbf{w}$ have same value of the phase.
The point $\bar\mathbf{w}$ is on the limit cycle, and, hence, its phase $\vp$
can be easily computed as described above, and $\vp(\mathbf{u})=
\vp(\mathbf{w}(0))=\vp(\bar\mathbf{w})=2\pi\frac{T-\tau(\bar\mathbf{w})}{T}$
is exactly the desired phase of the oscillator at the state $\mathbf{u}$.

In order to obtain the phase derivative we also have to follow the
autonomous evolution \eqref{app:eq2} towards the limit cycle
of the initial condition $\mathbf{u}+\mathbf{v}dt$.
(Practically, it can be performed simultaneously with
the evolution of the point $\mathbf{u}$.)
Since $dt$ is (infinitely) small, this
can be done by tracing the linear evolution of $\mathbf{v}dt$ to
$\bar\mathbf{v}dt$ within time interval $nT$.
The law of this linear evolution is
given by the Jacobian of the original equations~\eqref{app:eq2}.
Thus, two states $\mathbf{u}$ and $\mathbf{u}+\mathbf{v}dt$ of the coupled
system~(\ref{app:eq1})
map to two points $\bar\mathbf{w}$ and $\bar\mathbf{w}+\bar\mathbf{v}dt$
on the limit cycle of the autonomous system~(\ref{app:eq2}).
These points are characterized by the phases $\vp$ and $\vp+d\vp$, respectively.
On the other hand, let us consider the evolution of the autonomous system, from
the point $\bar\mathbf{w}$ to $\bar\mathbf{w}+\bar\mathbf{v}dt$.
This evolution occurs along the limit cycle of the system~(\ref{app:eq2})
within time interval $\overline{dt}$. (Notice that generally $\overline{dt}\ne dt$).
The evolution is governed by the flow on the cycle, i.e.
$\bar\mathbf{w}+\bar\mathbf{v}dt=\bar\mathbf{w}+\mathbf{f}(\bar\mathbf{w})\overline{dt}$,
what yields $\overline{dt}=dt\left(\bar\mathbf{v}\cdot \mathbf{f}(\bar\mathbf{w})\right)/
\| \mathbf{f}(\bar\mathbf{w})\|^2$.
Accordingly, phase growth is determined by the natural frequency $\w=2\pi/T$,
i.e. $d\vp=\w\overline{dt}$, which finally yields
\[
\frac{d\vp}{dt}=\w\frac{\bar\mathbf{v}\cdot \mathbf{f}(\bar\mathbf{w})}
{\|\mathbf{f}(\bar\mathbf{w})\|^2}\;.
\]

Thus, with the presented algorithm we can compute phases and their derivatives
as time series with an arbitrary time step and of sufficient length.
Certainly, this can be done for all elements of the network.

\subsection{Numerical reconstruction of the phase dynamics equations}
\label{sec:numerics}

Now, we discuss how the phase dynamics equations of a network can be
constructed numerically from given phases and their derivatives.
To explain the procedure, it
is convenient to denote the r.h.s. of Eq.~(\ref{eq:Q}) as
$Q_k(\vp_1,\vp_2,\ldots,\vp_N)$, where the subscript $k=1,2,\ldots, N$
labels the oscillator.
$Q_k$ are commonly referred to as the coupling functions.
Since each $Q_k$ is $2\pi$-periodic with respect to all its arguments,
we can write it as a multiple Fourier series
\begin{equation}
  \dot\vp_k = a_{k;\vec{0}}
    + \sum_{\vec{l} \neq \vec{0}} \left[
      a_{k; \vec{l}} \cos(\vec{\vp} \cdot \vec{l})
      + b_{k; \vec{l}} \sin(\vec{\vp} \cdot \vec{l})
    \right],
  \label{eq_fourier_full}
\end{equation}
where $\vec{l}=(l_1,l_2,\ldots,l_N)$ and $\vec{\vp}=(\vp_1,\vp_2,\ldots,\vp_N)$ are
$N$-dimensional vectors of integer indices and phases, respectively.
$\vec{\vp} \cdot \vec{l}=\sum_{j=1}^N l_j\vp_j $ denotes the scalar
product between the vectors of phases and the mode indices.
Notice the relation between the Fourier coefficients $a_{k; \vec{l}}$, $b_{k; \vec{l}}$,
and coefficients in
Eqs.~(\ref{eq:ph2ord1}-\ref{eq:ph2ord3}):
\[
a_{k; \vec{l}}=a^{(0)}_{k; \vec{l}} +\e a^{(1)}_{k; \vec{l}} +\e^2 a^{(2)}_{k; \vec{l}} \ldots\;,
\qquad b_{k; \vec{l}}= \e b^{(1)}_{k; \vec{l}} +\e^2 b^{(2)}_{k; \vec{l}} \ldots\;.
\]

Thus, we have time series of the phases
$\vp_{1,n},\vp_{2,n},\ldots,\vp_{N,n}$ and their derivatives $\dot\vp_{1,n},\dot\vp_{2,n},\ldots,\dot\vp_{N,n}$,
where $n=1,2,\ldots,L$ is the point index,
Eq.~(\ref{eq_fourier_full}) becomes a system of $L$ linear equations for
the unknown coupling coefficients $a,b$.
It is natural to approximate $Q_k$ by a finite Fourier series,
preserving only $M\ll L$ terms in the sum in Eq.~(\ref{eq_fourier_full}).
Practically, we vary the mode indices in the range $|l_j|\le m$, which leaves
$M = 2(m+1)(2m(2m+1)+1)-1$ unknown coefficients.
We apply the least square fit method to find the Fourier modes in the coupling
from the over-determined linear system. Practically
this is accomplished via the Singular Value Decomposition as
described in~\cite{press2007numerical}.

We denote the Fourier coefficients of the truncated series obtained from this procedure by
capitals $A_{k; \vec{l}},B_{k; \vec{l}}$. Thus, as a result of the
numerical evaluation we obtain an approximation of the phase dynamics as
\begin{equation}
  \dot\vp_k = A_{k;\vec{0}}
    + \sum_{\vec{l} \neq \vec{0}, |l_j|\le m} \left[
      A_{k; \vec{l}} \cos(\vec{\vp} \cdot \vec{l})
      + B_{k; \vec{l}} \sin(\vec{\vp} \cdot \vec{l})
    \right]\;.
  \label{eq_fourier_trunc}
\end{equation}
The sum in Eq.~(\ref{eq_fourier_full}) goes over the combination of mode indices
such that either $\vec{l}$ or $-\vec{l}$ is counted.

The success of this numerical approach depends on how how interdependent the series 
$\phi_k$ are.
Here we use two different protocols.

\paragraph{Asynchronous case.} Suppose that the network does not synchronize.
It means that the trajectories of the system   \eqref{eq:gen_network} are dense on the
$N$-dimensional torus.
Then one just has to calculate one long trajectory of
the system~(\ref{eq:gen_network}) starting from some initial
conditions and compute phases and instantaneous frequencies for each point of the
solution, as described above.
If the data series is sufficiently long, the computed trajectory covers the surface
of the torus spanned by $\vp,\vp_1,\vp_2,\ldots,\vp_N$.
Hence, we obtain enough information to recover the function $Q$ of $N$
variables and to determine the Fourier coefficients.

\paragraph{Synchronous case.}
If the network synchronizes, the trajectory on the torus collapses to
a closed line and Eqs.~\eqref{eq_fourier_full} for the Fourier coefficients
become dependent and cannot be solved.
However, there is a way to obtain enough data to solve the
problem even in this case. For this goal one considers not one long
trajectory, but an ensemble of short transients (cf.~\cite{Levnajic-Pikovsky-11,Pikovsky-18}).
Namely, one starts
numerical integration with some asynchronous initial conditions and follows
the trajectory unless it is on the invariant torus. The corresponding transient time
should be larger than the amplitude relaxation time, but smaller than the synchronization time of the phases.
Then the procedure is repeated with new initial conditions.
Thus, instead of one long record one collects many dynamical states on the torus,
unless the sufficient number of points is obtained.
Certainly, this protocol can be use for the asynchronous case as well.

We exemplify these two protocols in the next Section, but before proceeding
with the examples, we have to discuss an important issue.
As we mentioned, the least squares optimization requires that data points fill
the surface of an $N$-dimensional torus.
Thus, we face the curse of dimensionality: the data requirement grows
fast with the network size $N$ and becomes hardly feasible already for $N>3$.
Some information about the network, e.g., strength of directed links, can,
however, be revealed for $N>3$ as well.
A possible approach is to perform
the triplet-based analysis~\cite{kralemann2013detecting,kralemann2014reconstructing,osterhage2007measuring,rings2016distinguishing}.

\subsection{The Stuart-Landau network}
\label{sec: additive}

Here, we perform the numerical analysis for the system (\ref{eq:LS3})
with the goal to verify main results in Sec.~\ref{sec:3sltheory} as well
as the numerical approach. We choose two sets of parameters that correspond
to asynchronous and synchronous dynamics, respectively, and employ
the corresponding protocols.

The parameter values are:
$\a=0.1$, while $\w_1$, $\w_2$ and $\w_3$ are varied to change the
synchronization behaviour. Uncoupled oscillators with these parameters have a
limit cycle with $R = 1$ and initial conditions were chosen on it so that the
relaxation time is significantly reduced. The coupling parameters are all equal $c_{j,k}
= 1$ for $(j, k) \in \{(1,2), (2,1), (2,3), (3,2)\}$ and the phase lags are
$\b_{1,2} = 0.32$, $\b_{2,1} = 0.44$, $\b_{2,3} = 0.43$ and $\b_{3,2} = 0.18$.
We consider two sets of frequencies: case I with $\w_1 = -\sqrt{5}/2$, $\w_2 =
(\sqrt{2} - 1)/10$ and $\w_3 = 0.8$; here the dynamics is asynchronous
quasiperiodic; while for case II with $\w_1 = -0.055$, $\w_2 = 0$ and $\w_3 =
0.33$ the dynamics is synchronous with a long transient.
In both cases, we generated the set of points as follows:
we started the dynamics of system~\eqref{eq:LS3}
with random phases and amplitudes equal to one.
Then, after an initial transient time  $\Delta t =20$ (which has been chosen
to ensure that the relaxation of the amplitude to the invariant torus is over, but
locking of the phases still does not occur), the values of the phases and their velocities
were stored. Altogether we constructed a set of $L=10^6$ data points.

Next, we computed the coefficients of the truncated Fourier series, see
Eq.~(\ref{eq_fourier_trunc}), for different values of $\e$,
using the SVD approach as described above.
The system of coupled Stuart-Landau oscillators
 is invariant with respect to a phase
shift $\vp_k\to \vp_k+\phi$ which means that only  modes
that fulfill the condition
\begin{equation}
  \sum_{j=1}^3 l_j = 0 \label{eq slow condition}
\end{equation}
can exist.  To incorporate this into the analysis, we exploited two approaches:
\begin{enumerate}
\item[(i)] only modes satisfying \eqref{eq slow condition} are taken into account,
all other modes are set to zero; these modes constitute a small
subset of all possible modes and therefore we determined the Fourier coefficients
up to harmonics $|l_j|\le m=8$.
\item[(ii)] all modes with $|l_j|\le m=4$ are determined.
\end{enumerate}
Here we present the results for the case (i), while the results for (ii)
are presented in \ref{app:SLnum}.

\begin{figure}
\centering
  \includegraphics[width=0.7\textwidth]{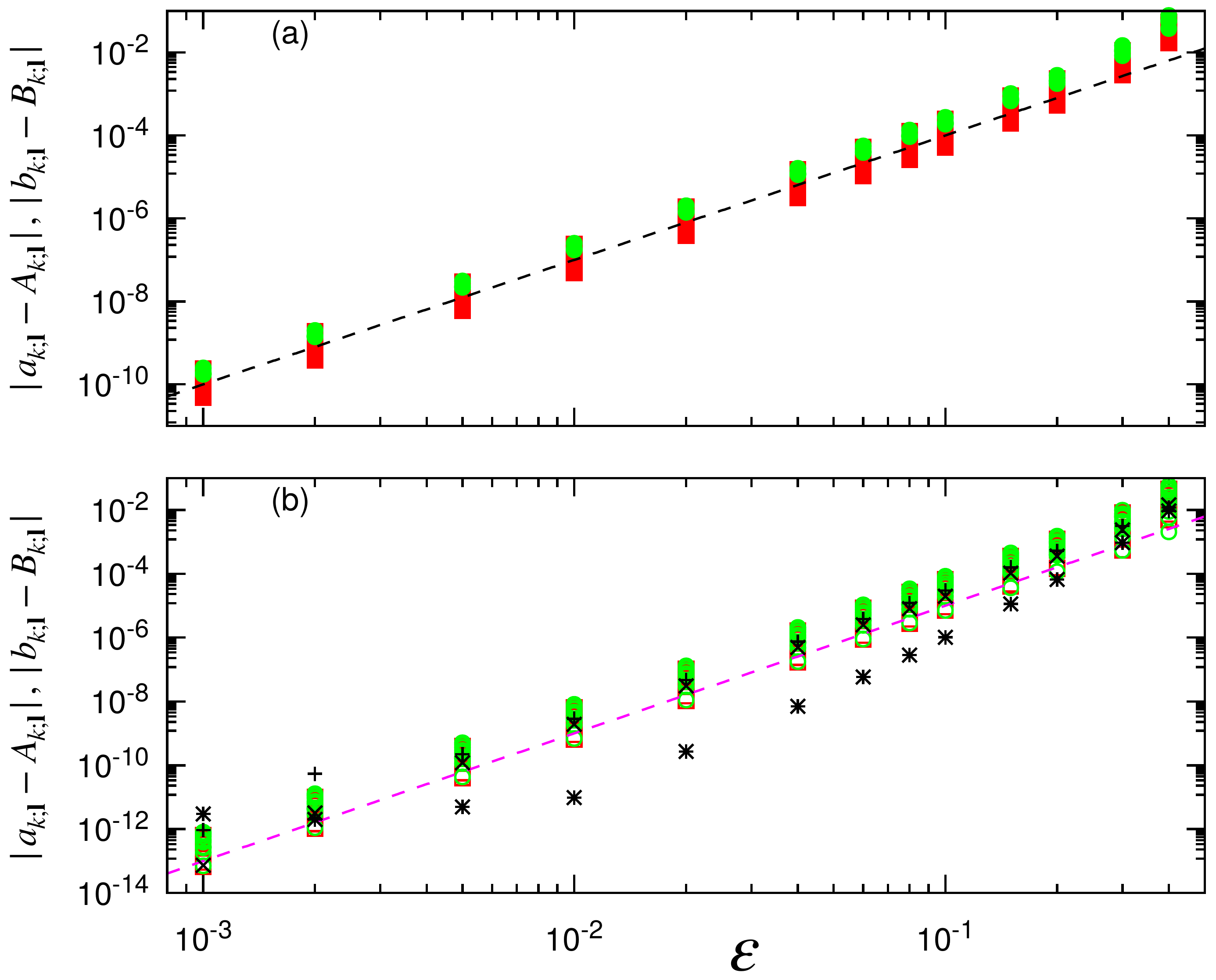}
  \caption{Results for the asynchronous configuration of three coupled
  Stuart-Landau oscillators, see Eq.~(\ref{eq:3Rph}).
  Differences between the numerically calculated coupling coefficients
  and the theoretical values given by Eqs.~(\ref{eq:ph2ord1}-\ref{eq:ph2ord3})
  are shown as a function of the coupling strength, for all oscillators.
  Panel (a) presents this difference for the terms appearing in the 1st order in $\e$;
  the black dashed line here corresponds to $\sim\e^ 3$.
  Panel (b) presents the terms appearing in the 2nd order in $\e$;
  the magenta dashed line here corresponds to $\sim\e^ 4$.
  Red squares (green circles) represent cosine (sine) coefficients.
  Additional black markers in (b) show
  the differences between the zero-order terms $A_{k;0}$, $k=1,2,3$
  and their theoretical values in the second-order approximation.}
  \label{fig:dif}
\end{figure}

\begin{figure}
\centering
  \includegraphics[width=0.7\textwidth]{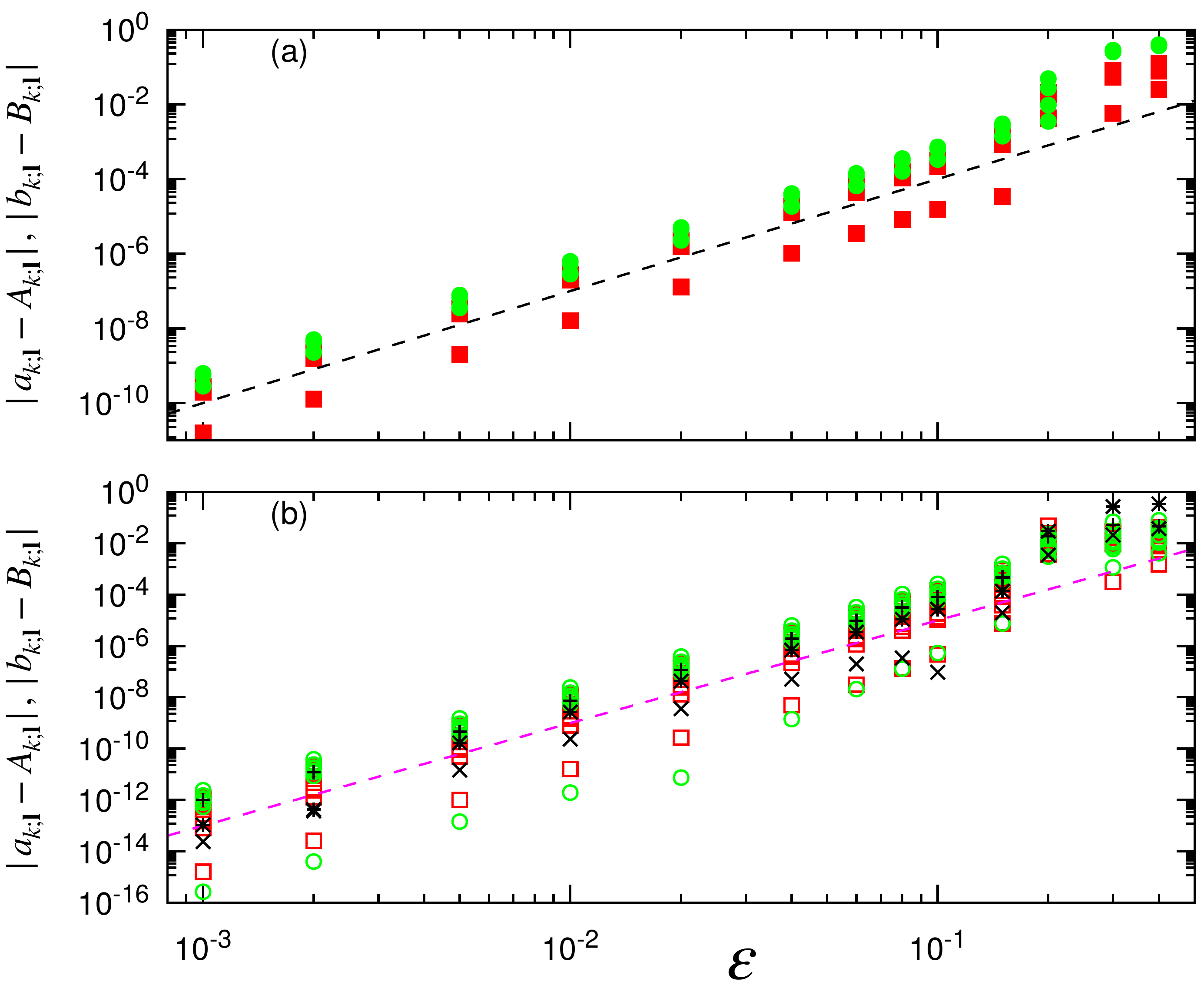}
  \caption{The same as in Fig.~\ref{fig:dif}, but for the synchronous configuration.}
  \label{fig:dif_s}
\end{figure}

First of all, we compare the theoretical findings presented
in Eqs.~(\ref{eq:ph2ord1}-\ref{eq:ph2ord3}) with numerical results.
To this end, we show in Figs.~\ref{fig:dif},\ref{fig:dif_s}  the
differences $|a_{k; \vec{l}}-A_{k; \vec{l}}|$, $|b_{k; \vec{l}}-B_{k; \vec{l}}|$,
where $k=1,2,3$, for the theoretically known modes
(see Eqs.~(\ref{eq:ph2ord1}-\ref{eq:ph2ord3})), for the asynchronous
and synchronous configurations, respectively.

We see that for weak coupling the difference is on the level of numerical precision;
it becomes of the order of one only for such strong coupling as $\e=0.4$.
Next, we see that the difference for the
first-order terms grows proportionally to $\e^3$, in correspondence
with our theoretical conclusion that there are no second-order correction
to the first-order terms. Thus, numerical results exhibit a good correspondence
with the theory.

\begin{figure}
\centering
  \includegraphics[width=0.7\textwidth]{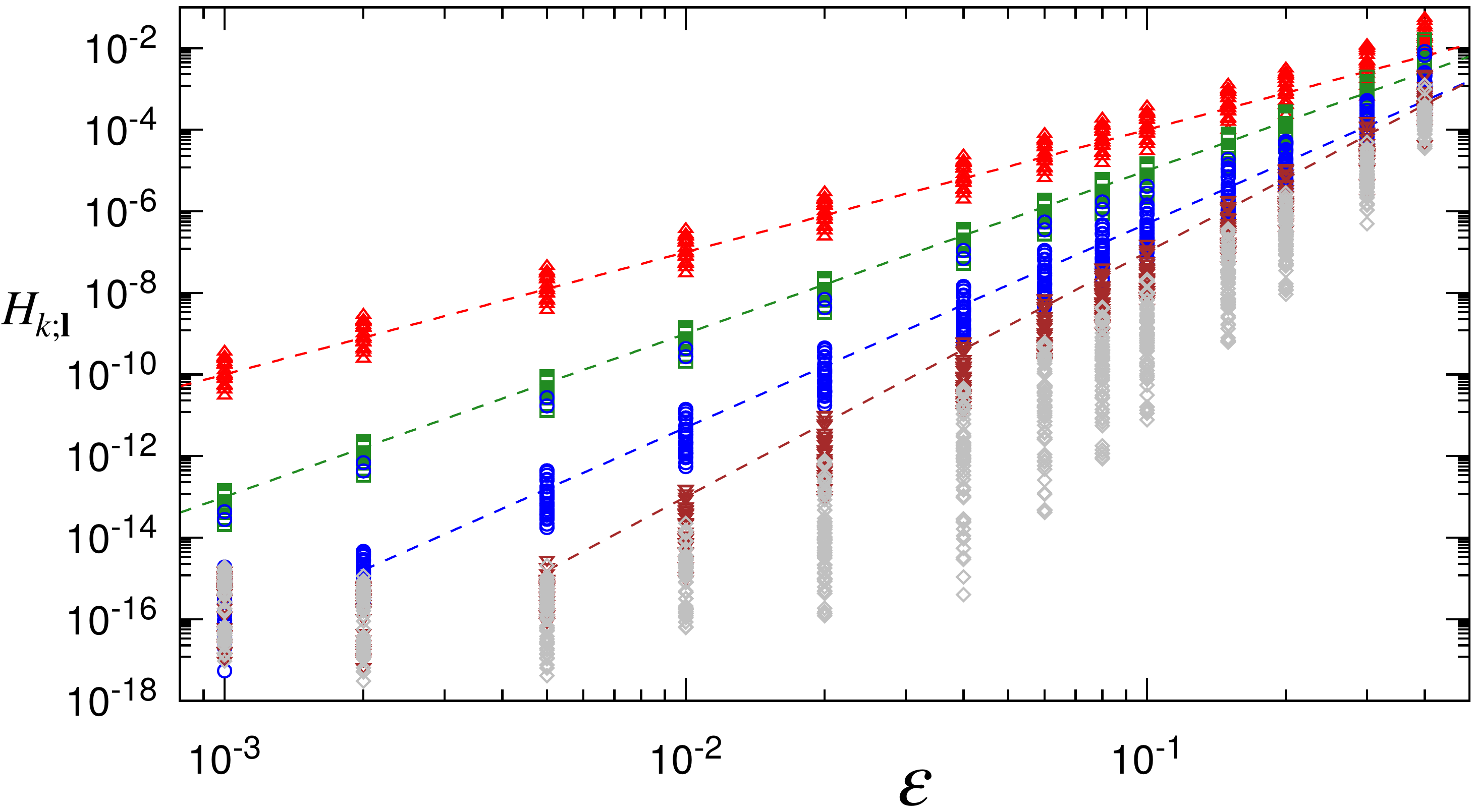}
  \caption{Results for the asynchronous configuration of three coupled
  Stuart-Landau oscillators, see Eq.~(\ref{eq:3Rph}). Here amplitudes
  $H_{k;\mathbf{l}}=\sqrt{A_{k;\mathbf{l}}^2+B_{k;\mathbf{l}}^2}$
  of all Fourier coefficients for all three oscillators
  except for those shown in Fig.~\ref{fig:dif} are plotted vs. the coupling strength.
  Red triangles up, green squares, blue circles, and brown triangles down show
  the coefficients that scale as $\e^3$, $\e^4$, $\e^5$, and $\e^6$, respectively.
  (The dashed lines, from top to bottom, have slopes 3, 4, 5, and 6 in
  log-log coordinates.)
  We did not checked for scaling $\sim\e^7$ and higher, and show all the coupling
  coefficients that do not fulfill above scaling laws with gray diamonds.
  }
  \label{fig:nontheory_modes}
\end{figure}

\begin{figure}
\centering
  \includegraphics[width=0.7\textwidth]{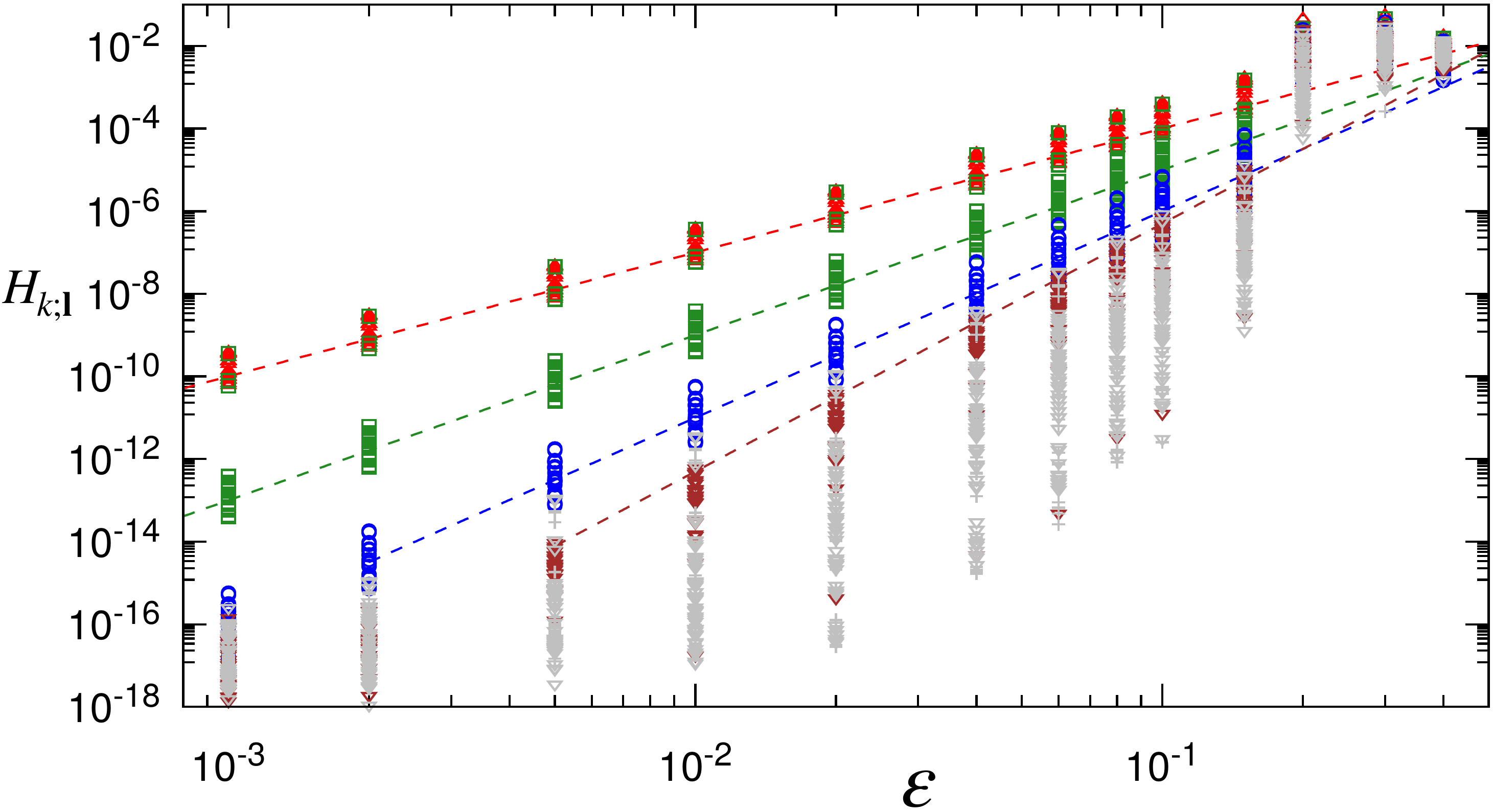}
  \caption{The same as in Fig.~\ref{fig:nontheory_modes}, but for the synchronous configuration.}
  \label{fig:nontheory_modes_s}
\end{figure}

Figures \ref{fig:nontheory_modes},\ref{fig:nontheory_modes_s} present an overview
of all Fourier coefficients for which we do not have theoretical values. Namely, we
show here all coefficients except for those entering
Eqs.~(\ref{eq:ph2ord1}-\ref{eq:ph2ord3}) and analyzed in
Figs.~\ref{fig:dif},\ref{fig:dif_s}.
The results are in full agreement with the power-series representation of the
coupling terms. Indeed, we see four groups of coefficients that scale as
$\e^3$, $\e^4$, $\e^5$ and $\e^6$, respectively.

\begin{figure}
\centering
  \includegraphics[width=0.5\textwidth]{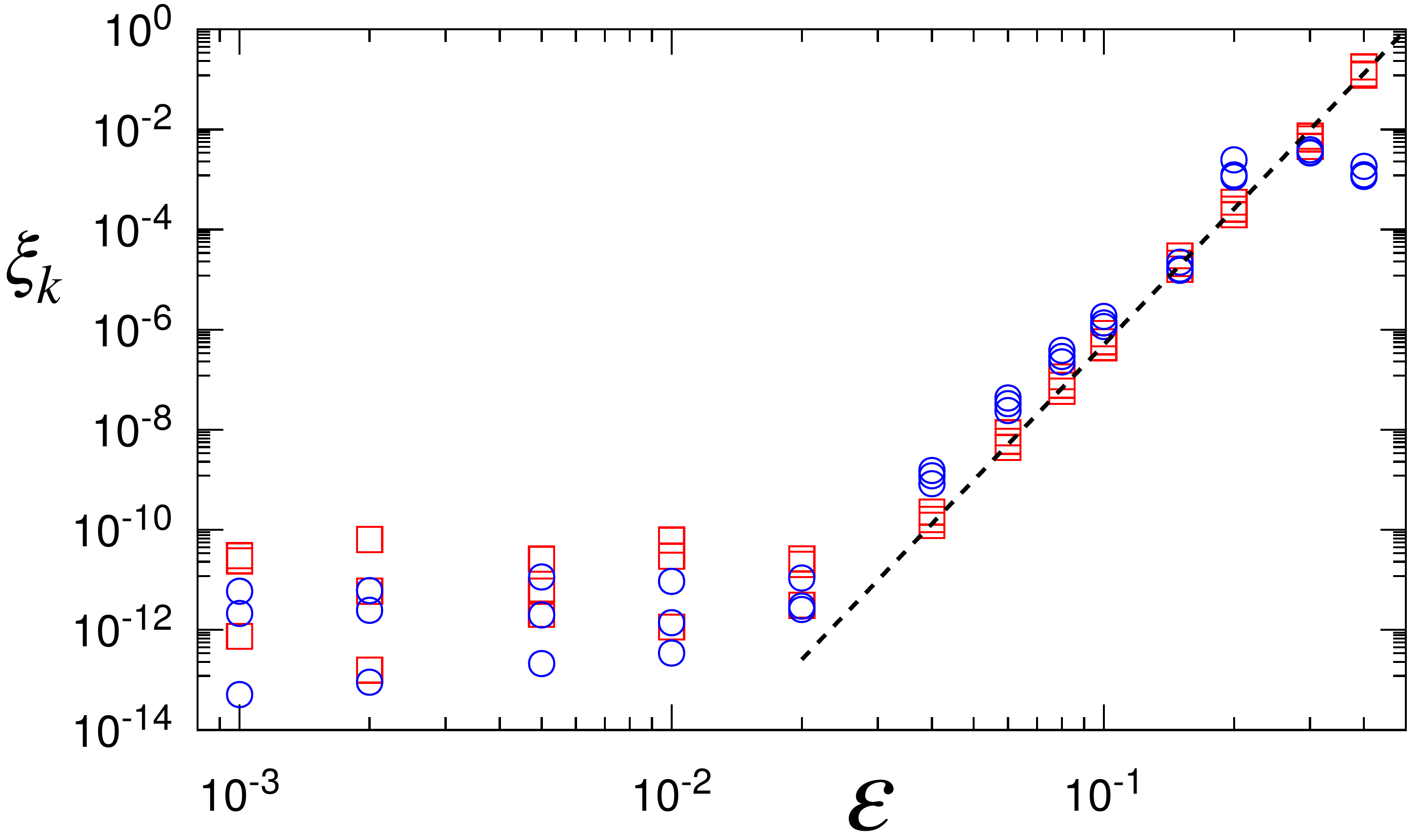}
  \caption{Accuracy of the phase dynamics reconstruction for all oscillators.
  Red squares and blue circles represent the rest term $\xi_k$ of the Fourier
  representation  Eq.~(\ref{eq_fourier_trunc}) for asynchronous and synchronous
  cases, respectively.
  The dashed line shows power law $\sim \e^9$.  }
  \label{fig:accur}
\end{figure}

Finally, we mention that the overall precision of the numerical procedure
can be estimated by computing the rest term $\xi_k$ of the Fourier series
representation in Eq.~(\ref{eq_fourier_trunc}).
The results presented
in Fig.~\ref{fig:accur} show that the rest term grows with the coupling strength
as $\e^9$. This is an indication that all
the terms of orders from $0$ to $8$ are within the set of chosen Fourier modes.

\subsection{A network of van der Pol oscillators}
Up to this point, our analysis was restricted to the case of Stuart-Landau
oscillators where expressions for phases and their derivatives are known.
In this Section we present a purely numerical evaluation of high-order
coupling terms for a network of
three non-identical van der Pol oscillators:
\begin{equation}
  \eqalign{
    \ddot x_1&-\mu(1-x_1^2)\dot x_1 +\omega_1^2 x_1 = \e x_2 \\
    \ddot x_2&-\mu(1-x_2^2)\dot x_2 +\omega_2^2 x_2 = \e (x_1+x_3) \\
    \ddot x_3&-\mu(1-x_3^2)\dot x_3 +\omega_3^2 x_3 = \e x_2 \\
  }
  \label{eq:vdp3}
\end{equation}
We fixed parameters $\omega_1=1$, $\omega_2= 1.324715957$, $\omega_3=\omega_2^2$
(here $\omega_2$
is the spiral mean, the root of cubic equation $\omega_2^3-\omega_2-1=0$).
We fixed $\mu=1$ and varied the coupling constant $\e$ in the range $0.001\leq \e\leq 0.3$, in this range the dynamics of the network is asynchronous.
From a trajectory of the system \eqref{eq:vdp3} we
obtained time series $\vp_{1,2,3},\dot\vp_{1,2,3}$ of length $L=10^6$ points each.
We used this set to estimate the coefficients of the truncated Fourier
series in Eq.~\eqref{eq_fourier_trunc} for $m=4$.
Thus, 729 coupling constants $A_{k;\vec{l}},\;B_{k;\vec{l}}$  were calculated
for each $\e$.
Below we restrict our attention only to the strengths of the coupling and ignore the relative phase, so we look on the properties of 364 coupling constants
$H_{k,\vec{l}}=(A_{k;\vec{l}}^2+B_{k;\vec{l}}^2)^{1/2}$. Together with the free term $A_{k;\vec{0}}$
this constitutes a set of $365$ numerically obtained coefficients for each oscillator and for each coupling constant $\e$.

For presentation of the results we perform a preliminary sorting.
According to the theory, we expect to obtain terms with leading dependencies
$\sim\e^q$,
with $q=0,1,2,\ldots$. Therefore, for each set of indices $\vec{l}$ and
each oscillator, we tried to approximate the coefficients $H_{k,\vec{l}}(\e)$ by the function $\sim \e^q$, and if the fitted value $q$ was close to an integer and the reliability of the fit was high, we attributed the corresponding power
to the coupling term.
Additionally, for $\e^0,\e,\e^2$ we used the five small values of $\e=0.001,\;0.002,\;0.005,\;0.01,\;0.02$, while
for powers $\e^3$ and $\e^4$ we used larger coupling constants $\e=0.04,\;0.06,\;0.08,\;0.1,\;0.15$. We did not look for powers $5$ and larger.

\begin{figure}
\centering
\includegraphics[width=0.6\textwidth]{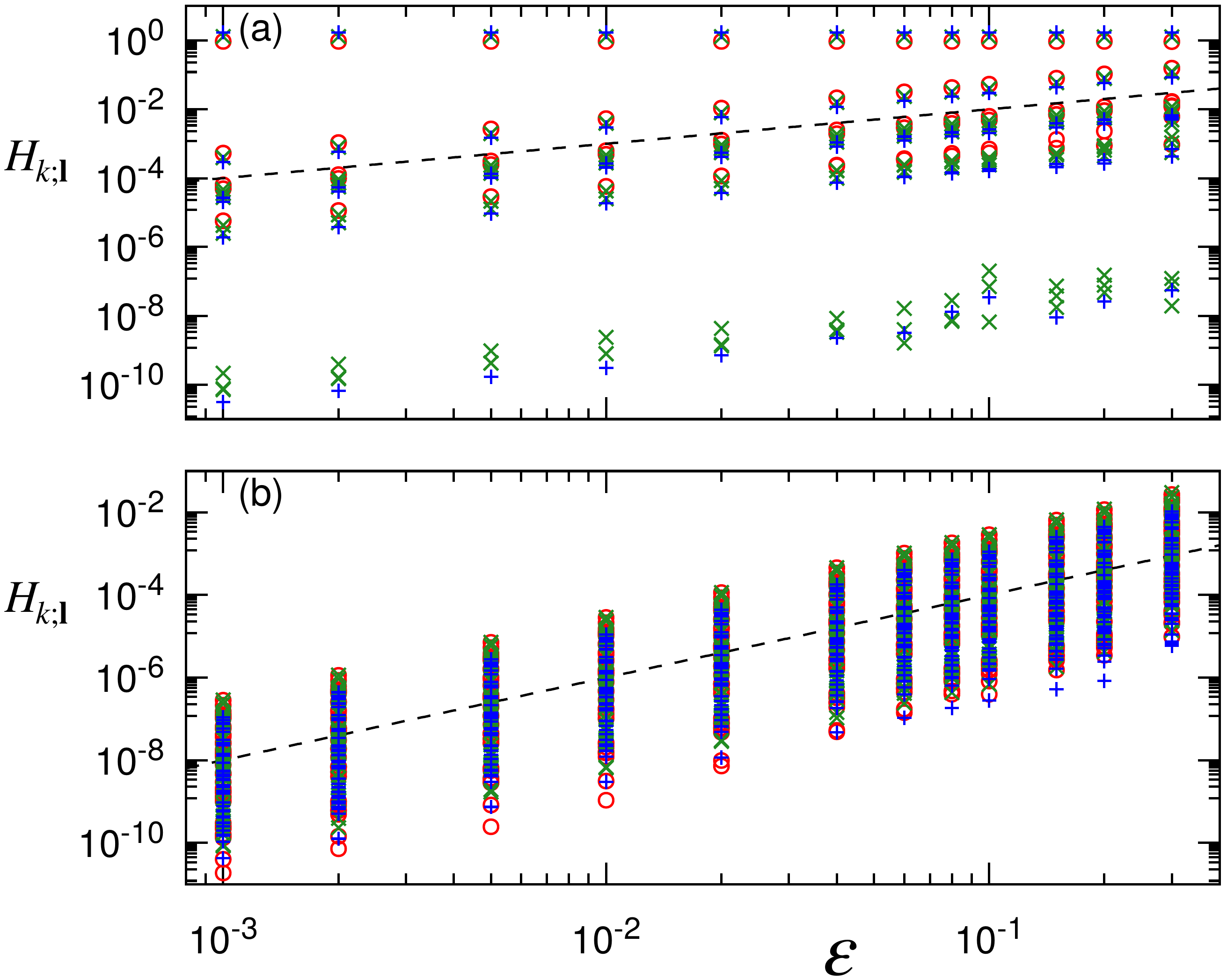}
\caption{Coupling coefficients $H_{k,\vec{l}}=(A_{k;\vec{l}}^2+B_{k;\vec{l}}^2)^{1/2}$
for all three oscillators are shown by red circles, green crosses,
and blue pluses, respectively, vs coupling strength $\e$.
Panel (a) shows powers $0$ and $1$, and panel (b) shows power $2$.
Dashed black lines correspond to the scaling $\sim\e$ and $\sim\e^2$,
respectively.
}
\label{fig:vdp1}
\end{figure}

Figure \ref{fig:vdp1} illustrates these findings.
It is instructive to look which coupling modes appear in each power of $\e$.
For modes up to $\sim\e^2$ we summarize this in Table~\ref{tab:vdpm}.
The modes $A_{k;\vec{0}}$,
describing corrections to the natural frequency are not listed.
Notice that in each cell of the table the modes are ordered according to their amplitudes.
Below we summarize the properties of the coupling modes of the van der Pol oscillators:
\begin{enumerate}
\item Looking at the modes that appear in the first order in $\e$, we notice
that the terms with phase differences and the terms with phase sums have
nearly the same amplitude.
This means that the coupling terms $\sim\e$ have nearly the Winfree form:
they are products of two functions of the oscillator phase and of the driving
oscillator phase, in full agreement with the first-order theory.
Recall that it is different for the Stuart-Landau oscillators, where only terms with
phase differences appear due to the condition~(\ref{eq slow condition}).
\item Because the variable $x$ of the van der Pol equation possesses odd harmonics
of the phase (and the same is true for the phase response curve),
terms with the third harmonics appear already in the 1st order in $\e$.
\item Inspection of terms $\sim\e$ in Table~\ref{tab:vdpm} reveals
some terms that are not expected
in the first-order analysis, we show them in italic font.
Data in Fig.~\ref{fig:vdp1} show that the amplitudes of
these terms are extremely small, comparable to the errors in terms reconstruction.
We conclude that these terms are probably spurious and just occasionally possess a scaling $\sim \e$, what is not surprising due to the fact that the total number of terms to be found is large.
\item With the size of the time series explored, we could not reliably
detect coupling terms appearing in order $\e^5$.
However, as the figures show, the terms $\sim 4$ can be detected
with good confidence.
\end{enumerate}

\begin{figure}
\centering
\includegraphics[width=0.6\textwidth]{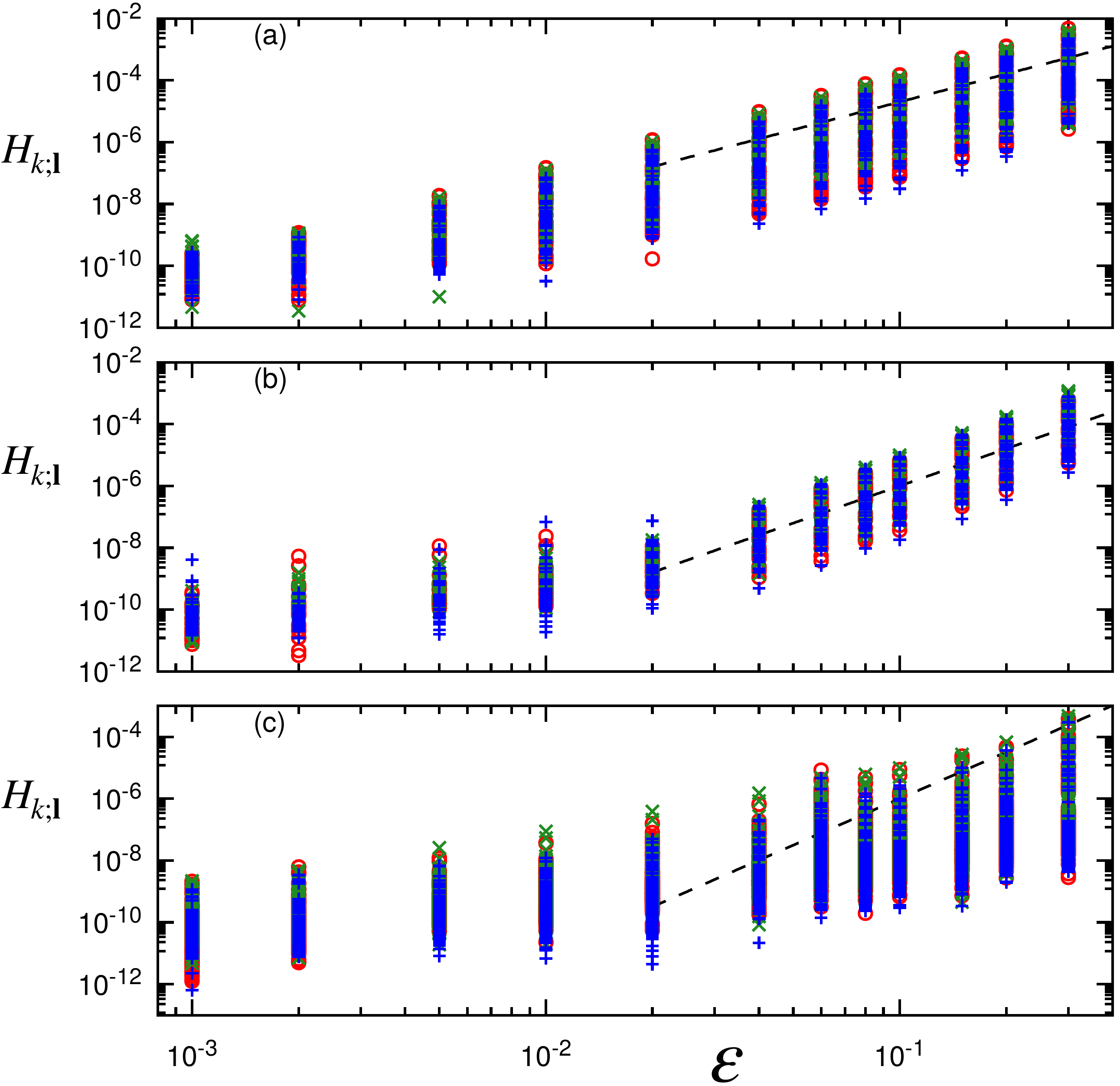}
\caption{The same as Fig.~\ref{fig:vdp1}, but for powers $3$ (a) and
$4$ (b). Panel (c) presents all other coefficients.
Dashed black lines show powers 3,4, and 5, respectively.
}
\label{fig:vdp2}
\end{figure}

\clearpage
\begin{table}
\begin{tabular}{|p{0.04\textwidth}|p{0.25\textwidth}|p{0.6\textwidth}|}
\hline
Osc & $\e$ & $\e^2$\\
\hline
1 & 8 Modes:  (1,-1,0), (1,1,0), (3,-1,0), (3,1,0), (1,3,0),  (1,-3,0), (3,-3,0), (3,3,0)
&
47 Modes: (2,-2,0),   (2,0,0),   (0,2,0),    (1,-2,1),   (1,2,-1),
(1,0,-1),    (1,0,1),   (2,2,0),    (1,-2,-1),    (1,2,1),    (2,-4,0),
(1,-4,1),    (1,4,-1),    (0,4,0),   (4,-2,0),    (2,4,0),    (3,-2,1),    (3,2,-1),
    (3,-2,-1),
(3,0,1),
(1,4,1),    (3,-4,1),    (4,2,0),    (4,-4,0),    (3,0,-1),    (1,-4,-1),    (3,-4,-1),
(4,0,0),    (3,2,1),    (3,4,-1),    (4,4,0),    (1,2,-3),    (1,0,-3),    (1,-2,3),
(3,4,1),    (1,0,3),    (1,2,3),    (1,-2,-3),    (3,-2,-3),    (3,-4,-3),    (3,-2,3),    (1,1,-4),    (3,0,-3),
    (1,4,3),    (3,2,-3),    (1,-4,-3),    (3,2,3) \\
\hline
2 & 19 Modes: (0,1,-1), (1,1,0), (0,1,1), (1,-1,0), (3,1,0), (3,-1,0), (0,3,1), (0,1,-3), (0,1,3), (1,3,0),
(1,-3,0), (3,-3,0), (0,3,-3), (3,3,0), (0,3,3), \textit{(1,2,0), (3,4,0), (2,1,4)}
& 56 Modes: (0,2,0), (2,-2,0), (1,-2,1), (0,2,-2), (2,0,0), (1,0,1), (1,0,-1), (1,-2,-1), (0,0,2), (1,2,-1), (1,2,1), (2,2,0),
(4,-2,0), (0,2,2), (4,0,0), (4,2,0), (0,2,-4), (3,0,1), (3,0,-1), (0,0,4), (1,-4,1), (1,-2,3), (3,2,-1), (1,2,-3), (1,4,-1),
(1,-2,-3), (1,0,-3), (0,4,-2), (0,2,4), (1,-4,-1),
(3,-2,1), (2,4,0), (3,-2,-1), (1,-4,3), (0,4,0), (1,4,1), (1,2,3), (1,0,3), (2,-4,0), (3,4,-1), (4,-4,0), (0,4,2), (3,2,1),
(3,-4,1), (3,0,-3), (3,-2,-3), (0,4,-4), (4,4,0), (1,4,-3), (3,-2,3), (3,0,3), (1,-4,-3), (0,4,4), (1,4,3), (3,-4,-3), (3,2,3)
\\
\hline
3&  9 Modes: (0,1,1), (0,1,-1), (0,3,1), (0,3,-1), (0,1,3), (0,1,-3), (0,3,-3), (0,3,3), \textit{(1,-1,1)}
& 50 Modes: (0,2,-2), (0,0,2), (1,0,-1), (1,0,1), (1,-2,-1), (1,-2,1), (0,2,0), (1,2,1), (1,2,-1), (0,2,2), (0,4,-2), (1,-4,-1)
, (1,-4,1), (0,4,0), (1,4,1), (1,4,-1), (0,4,2), (3,-2,-1), (3,-4,-1), (1,0,3), (3,-2,1), (3,-4,1), (1,0,-3), (1,-2,-3), (1,-2,3), (0,2,-4), (1,2,3),
(1,-4,3), (1,2,-3), (0,2,4)
, (3,2,1), (3,2,-1), (3,0,-1), (0,0,4), (3,0,1), (1,4,-3), (1,-4,-3), (1,4,3), (3,-2,-3), (0,4,-4), (3,4,1), (3,-2,3), (0,4,4),
(3,4,-1), (3,-4,3), (3,0,3), (3,2,3), (3,2,-3), (3,0,-3), (3,4,3)\\
\hline
\end{tabular}
\caption{All modes revealed in orders $\e$ and $\e^2$ for a network of van der Pol oscillators. Italic font in the second column denotes the modes that shall not appear in the
first-order approximation.}
\label{tab:vdpm}
\end{table}


\section{Conclusion}
\label{sec:concl}

In this work, we have presented an analytic perturbation approach, allowing the derivation of the equations for the phase dynamics for general networks of Stuart-Landau oscillators. We exemplified this framework with calculations for a particular system of three units.  We demonstrated explicitly that already in the second order in coupling strength,  there exist coupling terms that are not present in the structural coupling configuration. This result confirms a general statement that phase connectivity generally differs from the structural connectivity of a network. In particular, in higher-order approximation
we find triplet coupling terms, which are characteristic for a hypernetwork.

Since analytic derivations of the phase dynamics equations for generic oscillators remain a theoretical 
challenge, we developed a numerical method to compute the phase dynamics of an oscillatory network and to 
extract the coupling terms from time series of the obtained phases. These data could be one long trajectory if the
dynamics is quasiperiodic, or multiple small pieces (or points) if the dynamics is synchronous.
The result of the numerical procedure is a set of Fourier modes
of the coupling functions. Analysing these sets for different coupling strengths, we found terms with various power-law dependencies on this strength.
We first tested this approach on the coupled Stuart-Landau oscillators, where we demonstrated an excellent agreement with the theory.
This numerical approach has also been applied to a network of van der Pol oscillators, coupling functions of which are much more involved.

\ack
E.~Gengel acknowledges financial support from the Friedrich-Ebert-Stiftung.
A.~Pikovsky thanks Russian Science Foundation (Grant Number 17-12-01534) for support.
This paper was developed within the scope of the IRTG 1740 / TRP 2015/50122-0, funded by the DFG / FAPESP.

\appendix
\section{Second-order coupling coefficients in a network of Stuart-Landau oscillators.}

Here we present the coefficients for $\dot{\vp}_{1,2,3}$
in Eqs.~(\ref{eq:ph2ord1}-\ref{eq:ph2ord3}).
In Tables \ref{app tab 1}, \ref{app tab 2}, \ref{app tab 3} we use the notation
\[
C_{m,k}=\frac{1+\alpha^2}{2}\frac{2c_{m,k}}{4+(\w_m-\w_k)^2},\qquad D_{m,k}=\frac{1+\alpha^2}{2}\frac{(\w_m-\w_k)c_{m,k}}{4+(\w_m-\w_k)^2}\;.
\]

\begin{table}[!htbp!]
\centering
\begin{tabular}{|c|c|}
\hline
$a^{(2)}_{1;0}$ & $c_{2,1}\Big(C_{1,2}\sin(\beta_{1,2}+\beta_{2,1})
-D_{1,2}\cos(\beta_{1,2}+\beta_{2,1})-D_{2,1} \Big)$\\
$a^{(2)}_{1;-2,2,0}$ & $c_{2,1}\Big(C_{1,2}\sin(\beta_{2,1}-\beta_{1,2})+
D_{1,2}\cos(\beta_{2,1}-\beta_{1,2}) $\\
& $-C_{2,1}\sin 2\beta_{2,1}+D_{2,1}\cos2\beta_{2,1}\Big)$ \\
$b^{(2)}_{1;-2,2,0}$ & $c_{2,1}\Big(C_{1,2}\cos(\beta_{2,1}-\beta_{1,2})-
D_{1,2}\sin(\beta_{2,1}-\beta_{1,2})$ \\
& $-C_{2,1}\cos 2\beta_{2,1}-D_{2,1}\sin2\beta_{2,1}\Big)$ \\
$a^{(2)}_{1;-1,2,-1}$ & $c_{2,1}\Big(C_{3,2}\sin(\beta_{2,1}-\beta_{3,2})
+D_{3,2}\cos(\beta_{2,1}-\beta_{3,2})\Big)$ \\
$b^{(2)}_{1;-1,2,-1}$ & $c_{2,1}\Big(C_{3,2}\cos(\beta_{2,1}-\beta_{3,2})
-D_{3,2}\sin(\beta_{2,1}-\beta_{3,2})\Big)$ \\
$a^{(2)}_{1;-1,0,1}$ & $c_{2,1}\Big(-D_{3,2}\cos(\beta_{2,1}+\beta_{3,2})
+C_{3,2}\sin(\beta_{2,1}+\beta_{3,2})\Big)$ \\
$b^{(2)}_{1;-1,0,1}$ & $c_{2,1}\Big(D_{3,2}\sin(\beta_{2,1}+\beta_{3,2})
+C_{3,2}\cos(\beta_{2,1}+\beta_{3,2})\Big)$ \\
\hline
\end{tabular}
\caption{Coupling coefficients of the first Stuart-Landau oscillator.}
\label{app tab 1}
\end{table}

\begin{table}[!htbp!]
\centering
\begin{tabular}{|c|c|}
\hline
$a^{(2)}_{2;0}$ &  $\Big(C_{2,1}c_{1,2}\sin(\beta_{2,1}+\beta_{1,2})
-D_{2,1}c_{1,2}\cos(\beta_{2,1}+\beta_{1,2})-D_{1,2}c_{1,2}$ \\
& $+ C_{2,3} c_{3,2} \sin(\beta_{3,2}+\beta_{2,3}) - D_{3,2}c_{3,2} - D_{2,3} c_{3,2} \cos(\beta_{3,2}+\beta_{2,3})\Big)$ \\
$a^{(2)}_{2;2,-2,0}$ & $\Big(C_{2,1}c_{1,2}\sin(\beta_{1,2}-\beta_{2,1})+
D_{2,1}c_{1,2}\cos(\beta_{1,2}-\beta_{2,1})$\\
& $-C_{1,2}c_{1,2}\sin 2\beta_{1,2}+D_{1,2}c_{1,2} \cos2\beta_{1,2}\Big)$ \\
$b^{(2)}_{2;2,-2,0}$ & $\Big(C_{2,1}c_{1,2}\cos(\beta_{1,2}-\beta_{2,1})-
D_{2,1}c_{1,2}\sin(\beta_{1,2}-\beta_{2,1})$\\
& $-C_{1,2}c_{1,2}\cos 2\beta_{1,2} - D_{1,2}c_{1,2}\sin 2\beta_{1,2}\Big)$ \\
$a^{(2)}_{2;0,-2,2}$ & $\Big(C_{2,3}c_{3,2}\sin(\beta_{3,2}-\beta_{2,3})+
D_{2,3}c_{3,2}\cos(\beta_{3,2}-\beta_{2,3})$ \\
& $-C_{3,2}c_{3,2}\sin 2\beta_{3,2}+ D_{3,2}c_{3,2}\cos2\beta_{3,2}\Big)$ \\
$b^{(2)}_{2;0,-2,2}$ & $\Big(C_{2,3}c_{3,2}\cos(\beta_{3,2}-\beta_{2,3})-
D_{2,3}c_{3,2}\sin(\beta_{3,2}-\beta_{2,3})$ \\
& $-C_{3,2}c_{3,2}\cos 2\beta_{3,2} - D_{3,2}c_{3,2}\sin 2\beta_{3,2}\Big)$ \\
$a^{(2)}_{2;-1,2,-1}$ & $\Big(D_{3,2}c_{1,2}\cos(\beta_{1,2}+\beta_{3,2})
-C_{3,2}c_{1,2}\sin(\beta_{1,2}+\beta_{3,2})$ \\
& $ - C_{1,2} c_{3,2} \sin(\beta_{3,2}+\beta_{1,2}) + D_{1,2} c_{3,2} \cos(\beta_{3,2}+\beta_{1,2})\Big)$ \\
$b^{(2)}_{2;-1,2,-1}$ & $\Big(D_{3,2}c_{1,2}\sin(\beta_{1,2}+\beta_{3,2})
+C_{3,2}c_{1,2}\cos(\beta_{1,2}+\beta_{3,2}) $ \\
& $+ C_{1,2}c_{3,2}\cos(\beta_{3,2}+\beta_{1,2}) + D_{1,2}c_{3,2} \sin(\beta_{3,2}+\beta_{1,2}) \Big)$ \\
$a^{(2)}_{2;1,0,-1}$ & $\Big(-D_{3,2}c_{1,2}\cos(\beta_{1,2}-\beta_{3,2})
-C_{3,2}c_{1,2}\sin(\beta_{1,2}-\beta_{3,2})$ \\
& $ - C_{1,2}c_{3,2} \sin(\beta_{3,2}-\beta_{1,2}) - D_{1,2} c_{3,2} \cos(\beta_{3,2}-\beta_{1,2})\Big)$ \\
$b^{(2)}_{2;1,0,-1}$ & $\Big(D_{3,2}c_{1,2}\sin(\beta_{1,2}-\beta_{3,2})
-C_{3,2}c_{1,2}\cos(\beta_{1,2}-\beta_{3,2})$ \\
& $ + C_{1,2}c_{3,2} \cos(\beta_{3,2}-\beta_{1,2}) - D_{1,2}c_{3,2} \sin(\beta_{3,2}-\beta_{1,2})\Big)$ \\
\hline
\end{tabular}
\caption{Coupling coefficients of the second Stuart-Landau oscillator.}
\label{app tab 2}
\end{table}

\begin{table}[!htbp!]
\centering
\begin{tabular}{|c|c|}
\hline
$a^{(2)}_{3;0}$ & $c_{2,3}\Big(C_{3,2}\sin(\beta_{3,2}+\beta_{2,3})
-D_{3,2}\cos(\beta_{3,2}+\beta_{2,3})-D_{2,3} \Big)$\\
$a^{(2)}_{3;0,2,-2}$ & $c_{2,3}\Big(C_{3,2}\sin(\beta_{2,3}-\beta_{3,2})+
D_{3,2}\cos(\beta_{2,3}-\beta_{3,2})$ \\
& $-C_{2,3}\sin 2\beta_{2,3}+D_{2,3}\cos2\beta_{2,3}\Big)$ \\
$b^{(2)}_{3;0,2,-2}$ & $c_{2,3}\Big(C_{3,2}\cos(\beta_{2,3}-\beta_{3,2})-
D_{3,2}\sin(\beta_{2,3}-\beta_{3,2})$ \\
& $-C_{2,3}\cos 2\beta_{2,3}-D_{2,3}\sin2\beta_{2,3}\Big)$ \\
$a^{(2)}_{3;-1,2,-1}$ & $c_{2,3}\Big(C_{1,2}\sin(\beta_{2,3}-\beta_{1,2})
+D_{1,2}\cos(\beta_{2,3}-\beta_{1,2})\Big)$ \\
$b^{(2)}_{3;-1,2,-1}$ & $c_{2,3}\Big(C_{1,2}\cos(\beta_{2,3}-\beta_{1,2})
-D_{1,2}\sin(\beta_{2,3}-\beta_{1,2})\Big)$ \\
$a^{(2)}_{3;1,0,-1}$ & $c_{2,3}\Big(-D_{1,2}\cos(\beta_{2,3}+\beta_{1,2})
+C_{1,2}\sin(\beta_{2,3}+\beta_{1,2})\Big)$ \\
$b^{(2)}_{3;1,0,-1}$ & $c_{2,3}\Big(D_{1,2}\sin(\beta_{2,3}+\beta_{1,2})
+C_{1,2}\cos(\beta_{2,3}+\beta_{1,2})\Big)$ \\
\hline
\end{tabular}
\caption{Coupling coefficients of the third Stuart-Landau oscillator.}
\label{app tab 3}
\end{table}
\clearpage
\section{Terms in the higher orders for coupled Stuart-Landau oscillators}
In Table~\ref{tab:homodes} we present coupling modes appearing in higher orders in $\e$.
Namely, we just give the vectors $\vec{l}$ of these modes.
Up to order $4$ we checked all of them both analytically
and numerically; for order $5$ we present only numerical results.
\begin{table}[!htbp!]
\begin{tabular}{|c|c|c|c|
c|c|}
\hline
 & 1 and 3& 2 and 4& 3& 4& 5\\
 \hline
 1 & (-1,1,0) & (0,0,0) & (-3,3,0), (0,-1,1)    & (-4,4,0), (0,-2,2) &(0,3,-3), (5,-5,0)\\
   &          & (-2,2,0) & (-1,3,-2), (-2,3,-1) & (-2,0,2), (2,-4,2) & (2,-5,3), (3,-5,2)\\
   &          & (-1,0,1) & (-1,-1,2), (-2,1,1)  & (-1,-2,3), (3,-2,-1)&(3,-1,-2), (2,1,-3) \\
   &          & (1,-2,1) &                      & (-1,4,-3), (-3,4,-1) & (4,-5,1), (1,-5,4)\\
   &			&		&			&				&(4,-3,-1), (1,3,-4)\\
 \hline
 2 & (1,-1,0) & (0,0,0) & (-3,3,0), (0,3,-3)    & (-4,4,0), (0,4,-4) & (5,-5,0), (0,5,-5)\\
   & (0,-1,1) & (2,-2,0) & (-1,3,-2), (-2,3,-1) & (-2,0,2), (2,-4,2) & (3,-1,-2), (1,2,-3)\\
   &          & (0,-2,2) & (-1,-1,2), (-2,1,1)  & (-1,-2,3), (3,-2,-1) & (1,3,-4), (4,-3,-1)\\
   &          & (-1,0,1) &                      & (-1,4,-3), (-3,4,-1) & (4,-5,1), (1,-5,4)\\
   &          & (1,-2,1) &                      &  (2,-5,3), (3,-5,2)&(3,-5,2), (2,-5,3)\\
\hline
 3& (0,1,-1) & (0,0,0) & (0,3,-3), (1,-1,0)    & (0,4,-4), (2,-2,0)&(3,-3,0), (0,5,-5) \\
  &          & (0,2,-2) & (-1,3,-2), (-2,3,-1) & (2,0,-2), (2,-4,2) & (3,-5,2), (2,-5,3)\\
  &          & (1,0,-1) & (-1,-1,2), (-2,1,1)  & (-1,-2,3), (3,-2,-1) &(3,-1,-2), (2,1,-3) \\
  &          & (1,-2,1) &                      & (-1,4,-3), (-3,4,-1) &(1,-5,4), (4,-5,1)\\
  &			&		&			&				&(1,3,-4), (4,-3,-1)\\
\hline
\end{tabular}
\caption{Coupling terms that appear in different orders (see columns)
and for all three oscillators (see rows).}
\label{tab:homodes}
\end{table}
\section{Numerical reconstruction of coupling for Stuart-Landau oscillators: approach (ii)}
\label{app:SLnum}

\begin{figure}
\centering
  \includegraphics[width=0.6\textwidth]{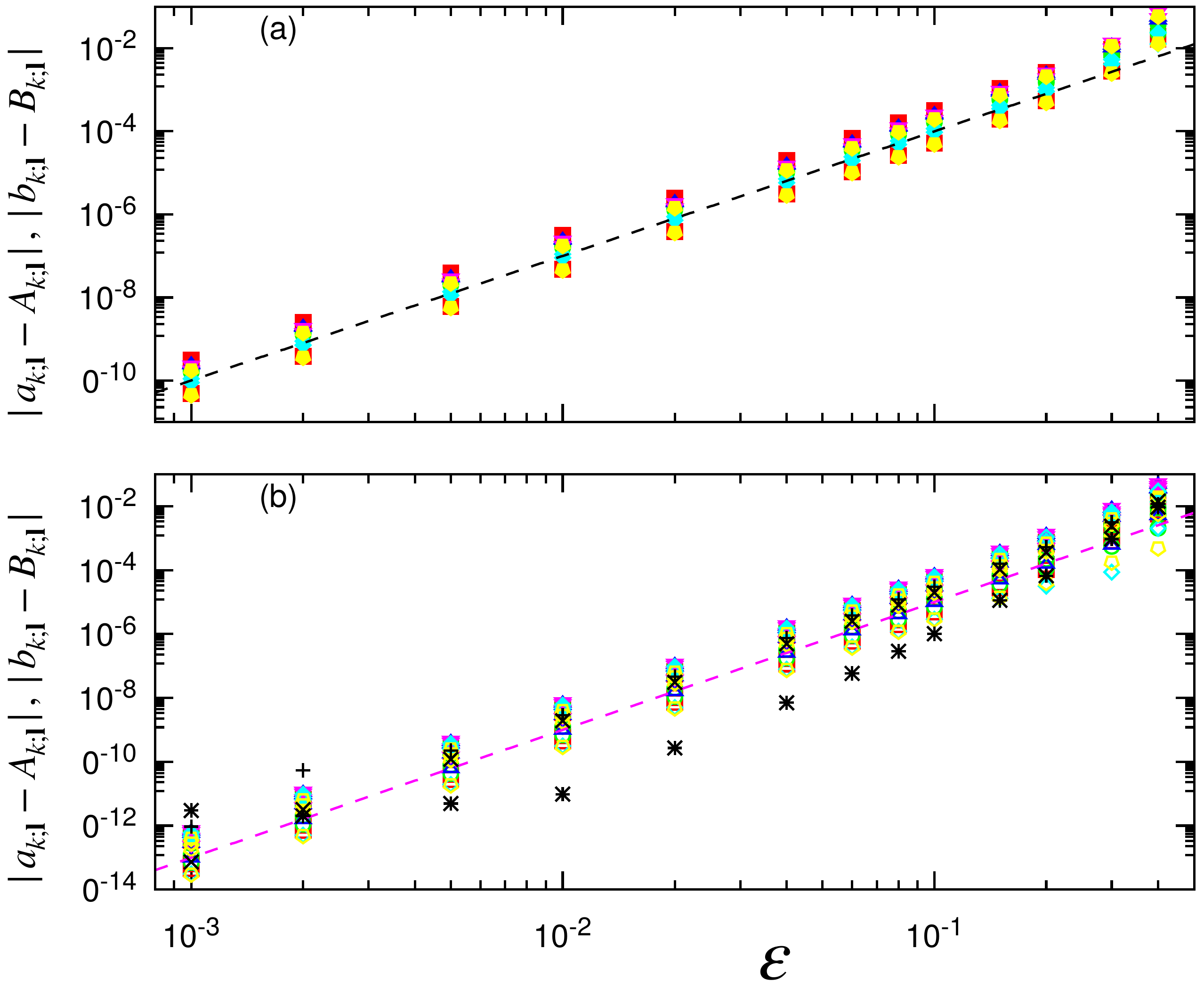}
  \caption{Results for the asynchronous configuration (case I) of three coupled
  Stuart-Landau oscillators, see Eq.(\ref{eq:3Rph}).
  Differences between the numerically calculated coupling coefficients
  and the theoretical values given by Eqs.(\ref{eq:ph2ord1}-\ref{eq:ph2ord3})
  are shown as a function of the coupling strength, for all oscillators.
  Panel (a) presents this difference for the terms appearing in the 1st order in $\e$;
  the black dashed line here corresponds to $\sim\e^ 3$.
  Panel (b) presents the terms appearing in the 2nd order in $\e$;
  the magenta dashed line here corresponds to $\sim\e^ 4$.
  Red squares (green circles) represent cosine (sine) coefficients.
  Additional black markers in (b) show
  the differences between the zero-order terms $A_{k;0}$, $k=1,2,3$
  and their theoretical values in the second-order approximation. }
  \label{fig:difx}
\end{figure}

\begin{figure}
\centering
  \includegraphics[width=0.5\textwidth]{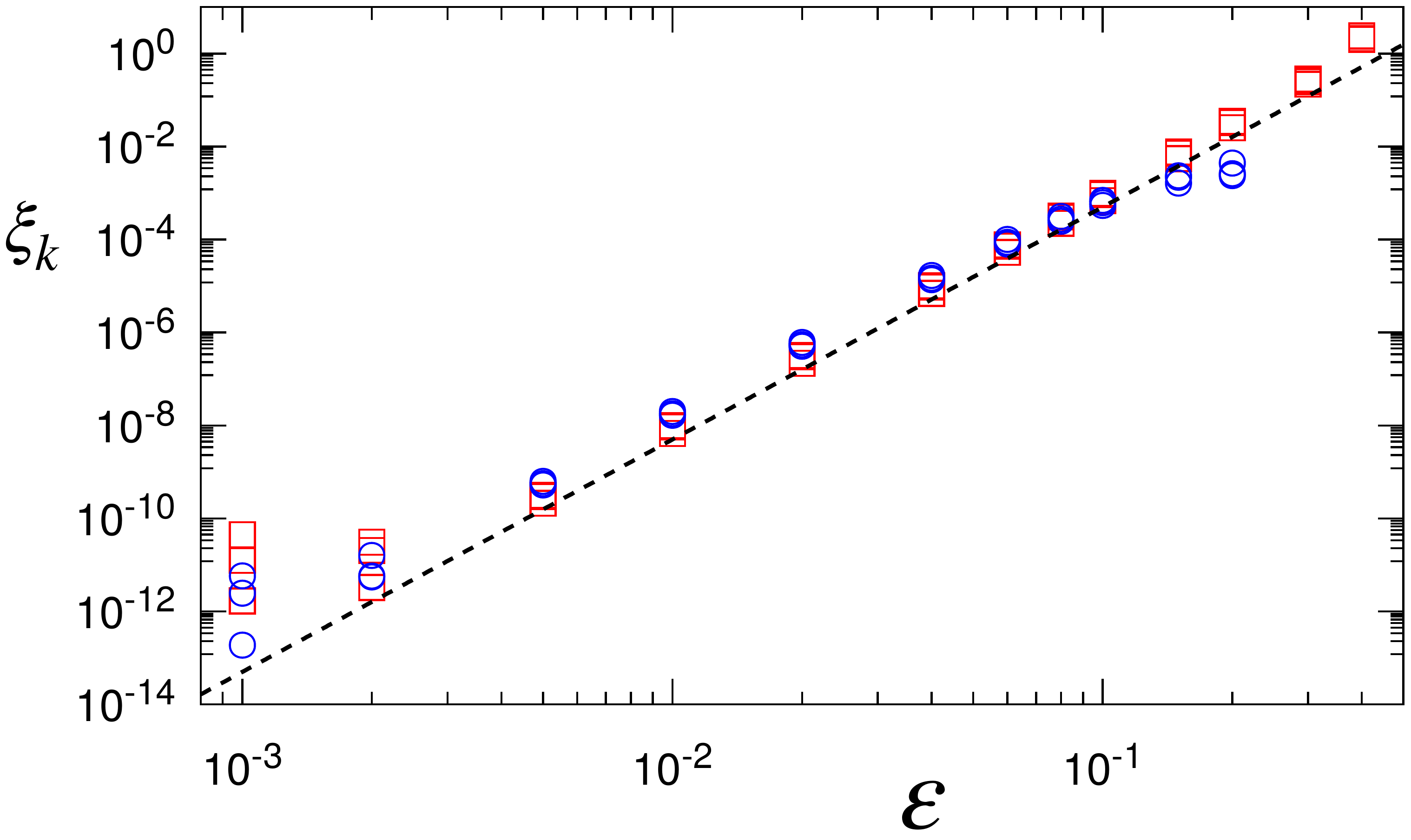}
  \caption{Accuracy of the phase dynamics reconstruction.
  Red squares and blue circles illustrate asynchronous and synchronous
  cases, respectively. The dashed line shows power law $\sim \e^5$.  }
  \label{fig:accurx}
\end{figure}

Here we present the result of the numerical procedure using the approach (ii), i.e.
here we do not exclude zero modes that do not fulfill the
condition (\ref{eq slow condition}).
Comparison of this case with the results for the approach (i),
presented in the main text, is important for the analysis
of networks of van der Pol or other oscillators, for which
no modes can be excluded \textit{a priori}.
Figures \ref{fig:difx}-%
\ref{fig:accurx} shall be compared to
Figs.~\ref{fig:dif}-%
\ref{fig:accur}, respectively.
The comparison shows that the results are consistent.
As expected, the approach (i) provides a higher accuracy since fewer unknowns
shall be found. However, even with the second approach, we managed to
reliably reveal scaling of the mode coefficients up to the order $\sim\e^5$.

\begin{figure}
\centering
\includegraphics[width=0.6\textwidth]{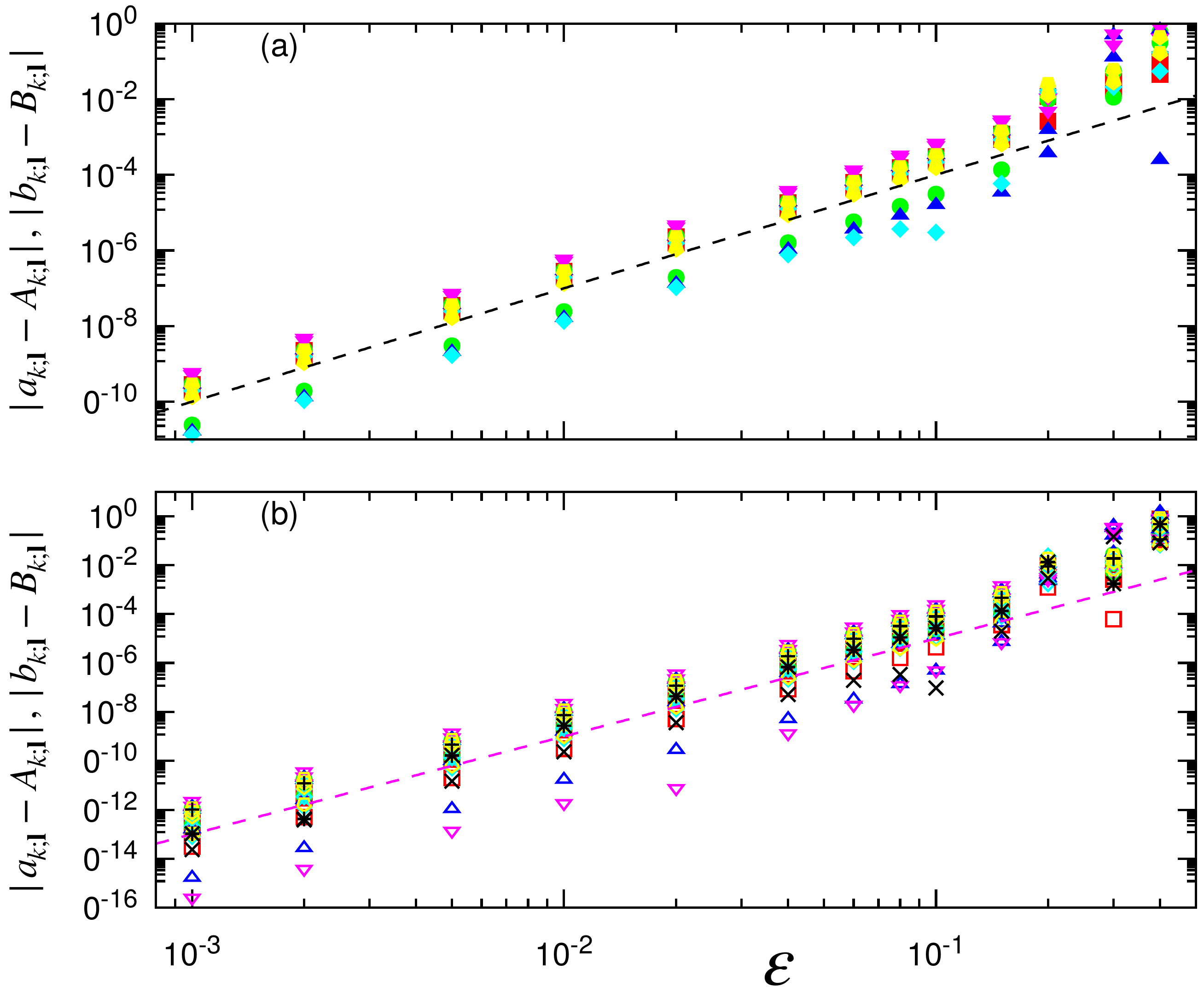}
  \caption{The same as in Fig.~\ref{fig:difx}, but for the synchronous
  configuration. }
  \label{fig:dif_sx}
\end{figure}

\begin{figure}
\centering
  \includegraphics[width=0.6\textwidth]{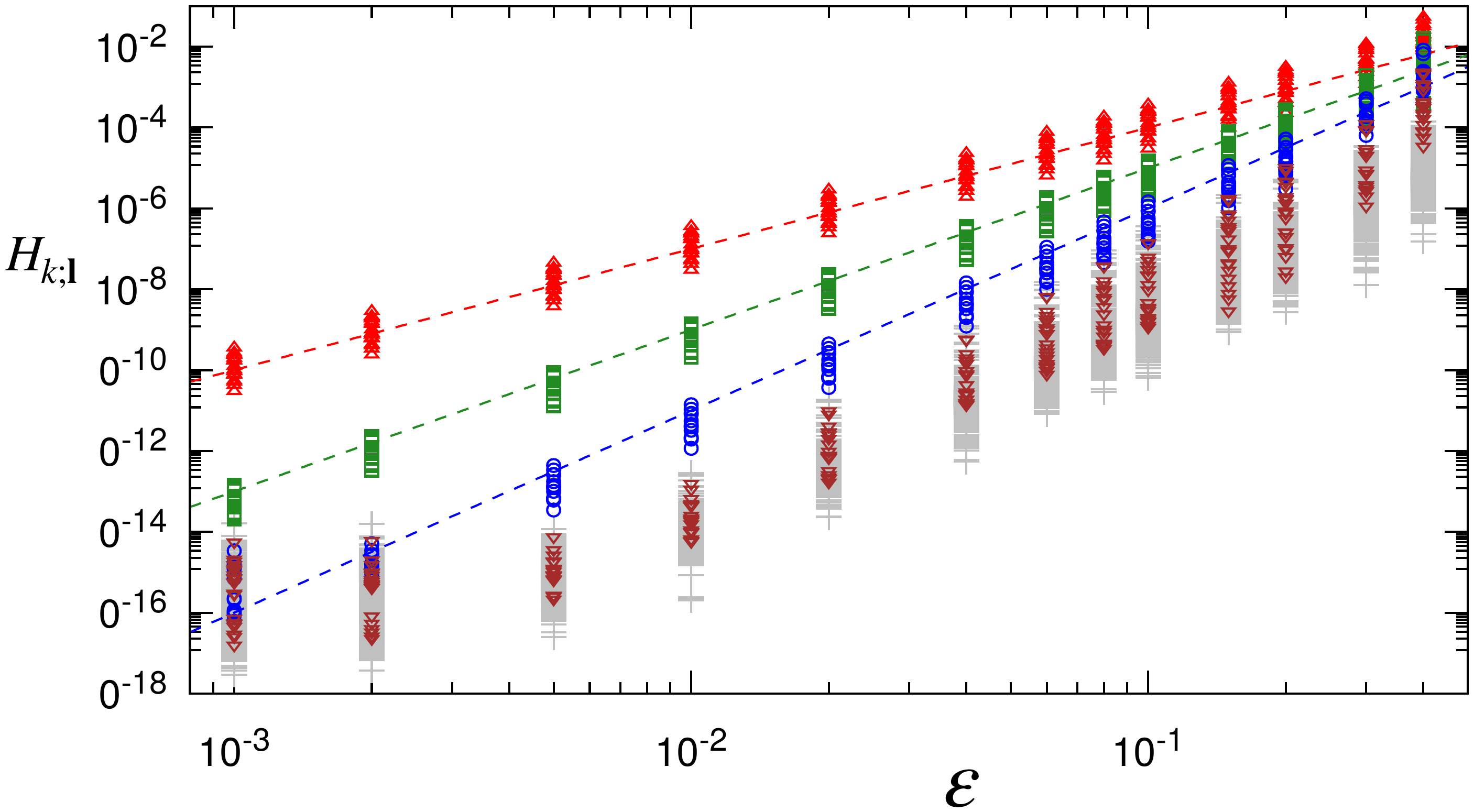}
  \caption{Results for the asynchronous configuration (case I)
  of three coupled
  Stuart-Landau oscillators, see Eq.~(\ref{eq:3Rph}). Here all Fourier coefficients
  for all three oscillators
  except for those shown in Fig.~\ref{fig:difx} are plotted vs. the coupling strength.
  Red triangles up, green squares, blue circles, and brown triangles down show
  the coefficients that scale as $\e^3$, $\e^4$, and $\e^5$, respectively.
  (The dashed lines, from top to bottom, have slopes 3, 4, and 5, in log-log coordinates.)
We did not checked for scaling $\sim\e^6$ and higher, and show all the coupling
coefficients that do not fulfill above scaling laws with
 gray pluses.
  }
  \label{fig:nontheory_modesx}
\end{figure}
\begin{figure}
\centering
  \includegraphics[width=0.6\textwidth]{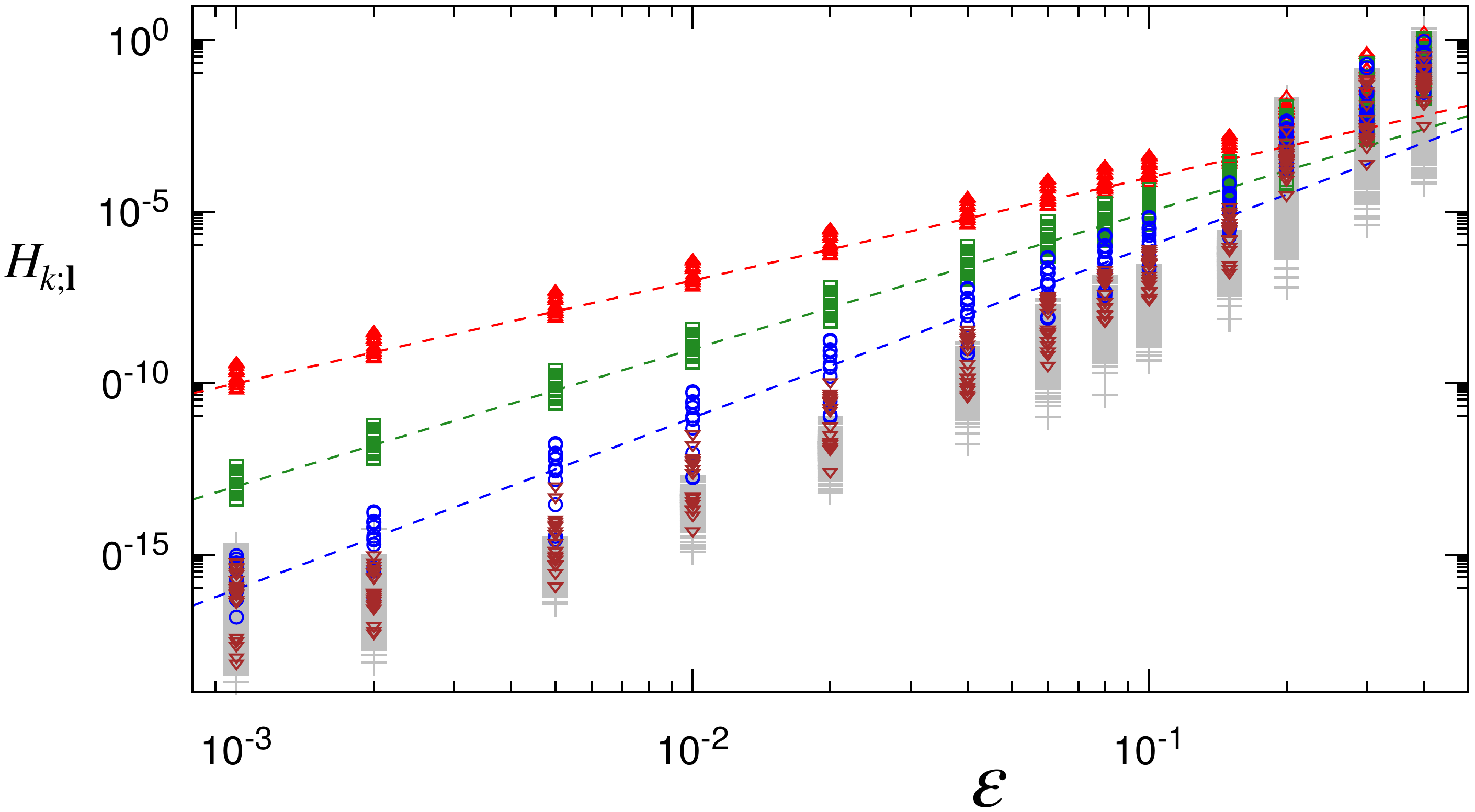}
  \caption{The same as in Fig.~\ref{fig:nontheory_modesx}, but for the synchronous configuration.}
  \label{fig:nontheory_modes_sx}
\end{figure}

\clearpage


\begin{thebibliography}{10}

\bibitem{Nixon_etal-13}
Micha Nixon, Eitan Ronen, Asher~A. Friesem, and Nir Davidson.
\newblock Observing geometric frustration with thousands of coupled lasers.
\newblock {\em Phys. Rev. Lett.}, 110:184102, 2013.

\bibitem{Matheny_etal-14}
M.~H. Matheny, M.~Grau, L.~G. Villanueva, R.~B. Karabalin, M.~C. Cross, and
  M.~L. Roukes.
\newblock Phase synchronization of two anharmonic nanomechanical oscillators.
\newblock {\em Phys. Rev. Lett.}, 112:014101, 2014.

\bibitem{pol1928heartbeat}
Balth. van~der Pol and J.~van~der Mark.
\newblock The heartbeat considered as a relaxation oscillation, and an
  electrical model of the heart.
\newblock {\em The London, Edinburgh, and Dublin Philosophical Magazine and
  Journal of Science}, 6(38):763--775, 1928.

\bibitem{ashraf2016synchronization}
I~Ashraf, R~Godoy-Diana, J~Halloy, B~Collignon, and B~Thiria.
\newblock Synchronization and collective swimming patterns in fish (hemigrammus
  bleheri).
\newblock {\em Journal of the Royal Society Interface}, 13(123):20160734, 2016.

\bibitem{strogatz2005theoretical}
Steven~H Strogatz, Daniel~M Abrams, Allan McRobie, Bruno Eckhardt, and Edward
  Ott.
\newblock Theoretical mechanics: Crowd synchrony on the {M}illennium {B}ridge.
\newblock {\em Nature}, 438(7064):43--44, 2005.

\bibitem{dorfler2014synchronization}
Florian D{\"o}rfler and Francesco Bullo.
\newblock Synchronization in complex networks of phase oscillators: A survey.
\newblock {\em Automatica}, 50(6):1539--1564, 2014.

\bibitem{Garca-Ojalvo99}
J.~Garc\'\i{}a-Ojalvo, J.~Casademont, and M.~C. Torrent.
\newblock Coherence and synchronization in diode-laser arrays with delayed
  global coupling.
\newblock {\em Int. J. of Bifurcation and Chaos}, 9(11):2225--2229, 1999.

\bibitem{Matheny_etal-19}
M.~H. Matheny, J.~Emenheiser, W.~Fon, A.~Chapman, A.~Salova, M.~Rohden, J.~Li,
  M.~Hudoba de~Badyn, M.~P{'o}sfai, L.~Duenas-Osorio, M.~Mesbahi, J.~P.
  Crutchfield, M.~C. Cross, R.~M. D’Souza, and M.~L. Roukes.
\newblock Exotic states in a simple network of nanoelectromechanical
  oscillators.
\newblock {\em Science}, 363:eaav7932, 2019.

\bibitem{Cawthorne_etal-99}
A.~B. Cawthorne, P.~Barbara, S.~V. Shitov, C.~J. Lobb, K.~Wiesenfeld, and
  A.~Zangwill.
\newblock Synchronized oscillations in {J}osephson junction arrays: The role of
  distributed coupling.
\newblock {\em Phys. Rev. B}, 60:7575--7578, 1999.

\bibitem{Tiberkevich_etal-09}
V.~Tiberkevich, A.~Slavin, E.~Bankowski, and G.~Gerhart.
\newblock Phase-locking and frustration in an array of nonlinear spin-torque
  nano-oscillators.
\newblock {\em Appl. Phys. Lett.}, 95:262505, 2009.

\bibitem{Motter-13}
A.~E. Motter, S.~A. Myers, M.~Anghel, and T.~Nishikawa.
\newblock Spontaneous synchrony in power-grid networks.
\newblock {\em Nat. Phys.}, 9:191, 2013.

\bibitem{Ulhaas_etal-09}
Peter Uhlhaas, Gordon Pipa, Bruss Lima, Lucia Melloni, Sergio Neuenschwander,
  Danko Nikolić, and Wolf Singer.
\newblock Neural synchrony in cortical networks: history, concept and current
  status.
\newblock {\em Frontiers in Integrative Neuroscience}, 3:17, 2009.

\bibitem{Glass-01}
L.~Glass.
\newblock {Synchronization and rhythmic processes in physiology}.
\newblock {\em Nature}, 410:277--284, 2001.

\bibitem{Prindle_etal-12}
A.~Prindle, P.~Samayoa, I.~Razinkov, T.~Danino, L.~S. Tsimring, and J.~Hasty.
\newblock A sensing array of radically coupled genetic ``biopixels''.
\newblock {\em Nature}, 481(7379):39--44, 2012.

\bibitem{kuramoto1984chemical}
Yoshiki Kuramoto.
\newblock {\em Chemical oscillations, turbulence and waves}.
\newblock Springer, Berlin, 1984.

\bibitem{pikovsky2003synchronization}
Arkady Pikovsky, Jurgen Kurths, Michael Rosenblum, and J{\"u}rgen Kurths.
\newblock {\em Synchronization: a universal concept in nonlinear sciences}.
\newblock Cambridge University Press, 2001.

\bibitem{Nakao-16}
Hiroya Nakao.
\newblock Phase reduction approach to synchronisation of nonlinear oscillators.
\newblock {\em Contemporary Physics}, 57(2):188--214, 2016.

\bibitem{pietras2019network}
Bastian Pietras and Andreas Daffertshofer.
\newblock Network dynamics of coupled oscillators and phase reduction
  techniques.
\newblock {\em Physics Reports}, 2019.

\bibitem{Kuramoto-75}
Y.~Kuramoto.
\newblock Self-entrainment of a population of coupled nonlinear oscillators.
\newblock In H.~Araki, editor, {\em International Symposium on Mathematical
  Problems in Theoretical Physics}, page 420, New York, 1975. Springer Lecture
  Notes Phys., v. 39.

\bibitem{Acebron-etal-05}
J.~A. Acebr{\'o}n, L.~L. Bonilla, C.~J.~P{\'e}rez Vicente, F.~Ritort, and
  R.~Spigler.
\newblock The {K}uramoto model: {A} simple paradigm for synchronization
  phenomena.
\newblock {\em Rev. Mod. Phys.}, 77(1):137--175, 2005.

\bibitem{leon2019phase}
Iv{\'a}n Le{\'o}n and Diego Paz{\'o}.
\newblock Phase reduction beyond the first order: The case of the mean-field
  complex {G}inzburg-{L}andau equation.
\newblock {\em Physical Review E}, 100(1):012211, 2019.

\bibitem{wilson2016isostable}
Dan Wilson and Jeff Moehlis.
\newblock Isostable reduction of periodic orbits.
\newblock {\em Physical Review E}, 94(5):052213, 2016.

\bibitem{daido1992order}
Hiroaki Daido.
\newblock Order function and macroscopic mutual entrainment in uniformly
  coupled limit-cycle oscillators.
\newblock {\em Progress of Theoretical Physics}, 88(6):1213--1218, 1992.

\bibitem{rosenblum2007self}
Michael Rosenblum and Arkady Pikovsky.
\newblock Self-organized quasiperiodicity in oscillator ensembles with global
  nonlinear coupling.
\newblock {\em Physical Review Letters}, 98(6):064101, 2007.

\bibitem{kurebayashi2013phase}
Wataru Kurebayashi, Sho Shirasaka, and Hiroya Nakao.
\newblock Phase reduction method for strongly perturbed limit cycle
  oscillators.
\newblock {\em Physical Review Letters}, 111(21), November 2013.

\bibitem{pyragas2015phase}
Kestutis Pyragas and Viktor Novi{\v{c}}enko.
\newblock Phase reduction of a limit cycle oscillator perturbed by a strong
  amplitude-modulated high-frequency force.
\newblock {\em Physical Review E}, 92(1), July 2015.

\bibitem{gengel2019phase}
Erik Gengel and Arkady Pikovsky.
\newblock Phase demodulation with iterative {H}ilbert transform embeddings.
\newblock {\em Signal Processing}, 165:115--127, 2019.

\bibitem{rosenblum2018inferring}
Rok Cestnik and Michael Rosenblum.
\newblock Inferring the phase response curve from observation of a continuously
  perturbed oscillator.
\newblock {\em Scientific reports}, 8(1):1--10, 2018.

\bibitem{kralemann2008phase}
Bj\"orn Kralemann, Laura Cimponeriu, Michael Rosenblum, Arkady Pikovsky, and
  Ralf Mrowka.
\newblock Phase dynamics of coupled oscillators reconstructed from data.
\newblock {\em Physical Review E}, 77(6), June 2008.

\bibitem{Bezruchko2003}
B.~Bezruchko, V.~Ponomarenko, M.~G. Rosenblum, and A.~S. Pikovsky.
\newblock Characterizing direction of coupling from experimental observations.
\newblock {\em CHAOS}, 13(1):179--184, 2003.

\bibitem{Tokuda-Jain-Kiss-Hudson-07}
I.~T. Tokuda, S.~Jain, I.~Z. Kiss, and J.~L. Hudson.
\newblock Inferring phase equations from multivariate time series.
\newblock {\em Phys. Rev. Lett.}, 99:064101, 2007.

\bibitem{Blaha-etal-11}
K.~A. Blaha, A.~Pikovsky, M.~Rosenblum, M.~T. Clark, C.~G. Rusin, and J.~L.
  Hudson.
\newblock Reconstruction of two-dimensional phase dynamics from experiments on
  coupled oscillators.
\newblock {\em Phys. Rev. E}, 84:046201, 2011.

\bibitem{Kralemann_etal-13}
B.~Kralemann, M.~Fr\"uhwirth, A.~Pikovsky, M.~Rosenblum, T.~Kenner,
  J.~Schaefer, and M.~Moser.
\newblock \textit{In vivo} cardiac phase response curve elucidates human
  respiratory heart rate variability.
\newblock {\em Nature Communications}, 4:2418, 2013.

\bibitem{Rosenblum-Fruehwirth-Moser-Pikovsky-19}
M.~Rosenblum, M.~Fr{\"u}hwirth, M.~Moser, and A.~Pikovsky.
\newblock Dynamical disentanglement in an analysis of oscillatory systems: an
  application to respiratory sinus arrhythmia.
\newblock {\em Philosophical Transactions of the Royal Society A: Mathematical,
  Physical and Engineering Sciences}, 377(2160):20190045, 2019.

\bibitem{ticcinelli2017coherence}
Valentina Ticcinelli, Tomislav Stankovski, Dmytro Iatsenko, Alan Bernjak,
  Adam~E Bradbury, Andrew~R Gallagher, Peter Clarkson, Peter~VE McClintock, and
  Aneta Stefanovska.
\newblock Coherence and coupling functions reveal microvascular impairment in
  treated hypertension.
\newblock {\em Frontiers in physiology}, 8:749, 2017.

\bibitem{topccu2018disentangling}
{\c{C}}a{\u{g}}da{\c{s}} Top{\c{c}}u, Matthias Fr{\"u}hwirth, Maximilian Moser,
  Michael Rosenblum, and Arkady Pikovsky.
\newblock Disentangling respiratory sinus arrhythmia in heart rate variability
  records.
\newblock {\em Physiological measurement}, 39(5):054002, 2018.

\bibitem{stankovski2017coupling}
Tomislav Stankovski, Tiago Pereira, Peter~VE McClintock, and Aneta Stefanovska.
\newblock Coupling functions: universal insights into dynamical interaction
  mechanisms.
\newblock {\em Reviews of Modern Physics}, 89(4):045001, 2017.

\bibitem{stankovski2016alterations}
Tomislav Stankovski, Spase Petkoski, Johan Raeder, Andrew~F Smith, Peter~VE
  McClintock, and Aneta Stefanovska.
\newblock Alterations in the coupling functions between cortical and
  cardio-respiratory oscillations due to anaesthesia with propofol and
  sevoflurane.
\newblock {\em Philosophical Transactions of the Royal Society A: Mathematical,
  Physical and Engineering Sciences}, 374(2067):20150186, 2016.

\bibitem{kralemann2011reconstructing}
Bj\"orn Kralemann, Arkady Pikovsky, and Michael Rosenblum.
\newblock Reconstructing phase dynamics of oscillator networks.
\newblock {\em Chaos: An Interdisciplinary Journal of Nonlinear Science},
  21(2):025104, 2011.

\bibitem{kralemann2014reconstructing}
Bj\"orn Kralemann, Arkady Pikovsky, and Michael Rosenblum.
\newblock Reconstructing effective phase connectivity of oscillator networks
  from observations.
\newblock {\em New Journal of Physics}, 16(8):085013, August 2014.

\bibitem{Komarov-Pikovsky-15b}
M.~Komarov and A.~Pikovsky.
\newblock Finite-size-induced transitions to synchrony in oscillator ensembles
  with nonlinear global coupling.
\newblock {\em Phys. Rev. E}, 92:020901, 2015.

\bibitem{Gong-Pikovsky-19}
C.~C. Gong and A.~Pikovsky.
\newblock Low-dimensional dynamics for higher-order harmonic, globally coupled
  phase-oscillator ensembles.
\newblock {\em Phys. Rev. E}, 100:062210, 2019.

\bibitem{rosenblum2019numerical}
Michael Rosenblum and Arkady Pikovsky.
\newblock Numerical phase reduction beyond the first order approximation.
\newblock {\em Chaos: An Interdisciplinary Journal of Nonlinear Science},
  29(1):011105, 2019.

\bibitem{press2007numerical}
William~H Press, Saul~A Teukolsky, William~T Vetterling, and Brian~P Flannery.
\newblock {\em Numerical recipes: The art of scientific computing (3rd
  edition)}.
\newblock Cambridge university press, 2007.

\bibitem{Levnajic-Pikovsky-11}
Z.~Levnaji\ifmmode~\acute{c}\else \'{c}\fi{} and A.~Pikovsky.
\newblock Network reconstruction from random phase resetting.
\newblock {\em Phys. Rev. Lett.}, 107(3):034101, 2011.

\bibitem{Pikovsky-18}
A.~Pikovsky.
\newblock Reconstruction of a random phase dynamics network from observations.
\newblock {\em Physics Letters A}, 382(4):147 -- 152, 2018.

\bibitem{kralemann2013detecting}
Bj\"orn Kralemann, Arkady Pikovsky, and Michael Rosenblum.
\newblock Detecting triplet locking by triplet synchronization indices.
\newblock {\em Physical Review E}, 87(5), May 2013.

\bibitem{osterhage2007measuring}
Hannes Osterhage, Florian Mormann, Tobias Wagner, and Klaus Lehnertz.
\newblock Measuring the directionality of coupling: phase versus state space
  dynamics and application to {EEG} time series.
\newblock {\em International journal of neural systems}, 17(03):139--148, 2007.

\bibitem{rings2016distinguishing}
Thorsten Rings and Klaus Lehnertz.
\newblock Distinguishing between direct and indirect directional couplings in
  large oscillator networks: Partial or non-partial phase analyses?
\newblock {\em Chaos: An Interdisciplinary Journal of Nonlinear Science},
  26(9):093106, September 2016.

\end{thebibliography}

\end{document}